\documentclass[a4paper,twocolumn,fleqn,usenatbib]{mnras}
\usepackage{mathptmx}

\usepackage[T1]{fontenc}
\usepackage{ae,aecompl}

\usepackage{graphicx}	
\usepackage{amssymb}
\usepackage{amsmath}
\usepackage{pdflscape}

\title[He-like triplet diagnostics]{A benchmark of the He-like triplet for ions with $6\leq Z\leq 14$ in Maxwellian and non-Maxwellian plasmas}

\author[Y. Rodr\'iguez et al.]{
Y. Rodr\'iguez,$^{1}$
E. Gatuzz,$^{1,2}$
M. A. Bautista,$^{3}$
C. Mendoza$^{4}$\thanks{E-mail: claudio@ivic.gob.ve}
\\
$^{1}$Escuela de F\'isica, Facultad de Ciencias, Universidad Central de Venezuela, A.P. 20513, Caracas 1020A, Venezuela\\
$^{2}$Max-Planck-Institut f\"ur Astrophysik, 85741, Garching bei M\"unchen, Germany\\
$^{3}$Department of Physics, Western Michigan University, 1903 W Michigan Ave., Kalamazoo, MI 49008, USA\\
$^{4}$Centro de F\'isica, Instituto Venezolano de Investigaci\'on Cient\'ificas (IVIC), A.P. 21827, Caracas 1020A, Venezuela
}

\date{Accepted XXX. Received YYY; in original form ZZZ}

\pubyear{2015}

\begin{document}
\label{firstpage}
\pagerange{\pageref{firstpage}--\pageref{lastpage}}
\maketitle

\begin{abstract}
  After an extensive assessment of the effective collision strengths available to model the He-like triplet of \ion{C}{v}, \ion{N}{vi}, \ion{O}{vii}, \ion{Ne}{ix}, \ion{Mg}{xi} and \ion{Si}{xiii} in collisionally dominated plasmas, new accurate effective collision strengths are reported for \ion{Ne}{ix}. The uncertainty intervals of the density and temperature diagnostics due to the atomic data errors are also determined for both Maxwell--Boltzmann and $\kappa$ electron-energy distributions. It is shown that these uncertainty bands limit the temperature range where the temperature line-ratio diagnostic can be applied and its effectiveness to discern the electron-energy distribution type. These findings are benchmarked with {\em Chandra} and {\em XMM-Newton} spectra of stellar coronae and with tokamak measurements.
\end{abstract}




\section{Introduction}

Since they were first proposed by \citet{gab69}, plasma diagnostics based on the three emission lines arising from the $n=2$ complex of He-like ions have been widely used in high-resolution X-ray spectroscopy to characterise both collisional and photoionised plasmas \citep{por00a, por00b, por01a, por01b, mew03, por05}. Regarding collisionally dominated plasmas, the main concern of the present report, {\em Chandra} and {\em XMM-Newton} spectra of stellar coronae displaying He-like emission triplets from \ion{C}{v}, \ion{N}{vi}, \ion{O}{vii}, \ion{Ne}{ix}, \ion{Mg}{xi} and \ion{Si}{xiii} have been useful in the study of stellar magnetic activity, coronal size, magnetic loop length scales, abundance anomalies, filling factors and the stellar radiation field \citep{nes01, nes02a, nes02b, nes04, tes04a, tes15, nes05b, leu07}.

The significantly higher spectral resolution and sensitivity that have been specified for the next generation X-ray space observatories will boost the diagnostic potential of the He-like triplet. For instance, it will be possible to resolve its resonance, intercombination and forbidden components in medium- to high-$Z$ ions (atomic number $Z\leq 30$) to characterise high-temperature plasmas and the satellite lines to test departures from ionisation equilibrium and Maxwellian electron-energy distributions \citep{por10}. In this respect, \citet{dzi00} have shown that deviations from a Maxwellian distribution in the solar corona affect the line intensities (e.g. of \ion{Fe}{xxv}) in a manner that suffices to diagnose flares; and more recently, \citet{dud15} have reported the first reliable evidence of a $\kappa$ distribution in a transient coronal loop. Important limitations in these endeavours are the uncertainties of both the observed line intensities and atomic databases \citep{dzi08}; in particular of reliable $\kappa$-averaged effective collision strengths that are derived from integrations of the energy tabulations of the collision strengths that are seldom published.

We are interested in evaluating the influence of the $\kappa$ distribution on the He-like triplet diagnostics of low- to medium-$Z$ ($Z\leq 14$) ions in coronal (collisionally dominated) plasmas, and in the possibilities of detecting deviations from the Maxwell-Boltzmann (MB) distribution once the uncertainties of the atomic data and observations are taken into account. For this purpose we recall in Section~\ref{pm} the expressions for the electron-impact effective collision strength in both MB and $\kappa$ distributions, and compute $\kappa$-averaged effective collision strengths from the collision-strength energy tabulations. The formalism of \citet{bau13} to estimate the emissivity-ratio uncertainty as propagated from the atomic data errors is extended to consider both MB and $\kappa$ distributions (Section~\ref{error_prog}). We make extensive tests to determine the accuracy of the available effective collision strengths for the aforementioned ions (Section~\ref{atomic_data}), and derive emissivity-ratio uncertainty bands for both MB and $\kappa$ distributions in Sections~\ref{MB}--\ref{kappa}. As a result, diagnostic maps are used in Section~\ref{obs} to infer plasma conditions from the observed spectra of several stellar coronae, and in Section~\ref{tokamak} some of the line ratios are compared with tokamak measurements. Finally, the main conclusions of the present benchmark are discussed in Section~\ref{conclusions}.

\section{He-like triplet diagnostics}

The He-like triplet is composed of three emission lines observed in members of the He isoelectronic sequence arising from transitions between excited levels with principal quantum number $n=2$ and the ${\rm 1\,^1S_0}$ ground state: the ${\rm 2\,^1P^o_1 -  1\,^1S_0}$ resonance line (referred to as $w$ or $r$); the intercombination line ($i=x+y$ where $x$ and $y$ correspond to the magnetic quadrupole ${\rm 2\,^3P^o_{2} - 1\,^1S_0}$ transition and the allowed spin-forbidden ${\rm 2\,^3P^o_{1}- 1\,^1S_0}$ transition, respectively) and the magnetic dipole ${\rm 2\,^3S_1 - 1\,^1S_0}$ forbidden line ($z$ or $f$). Spectral triplets from different cosmic abundant He-like ions are observed in a variety of astronomical plasmas such as stellar coronae, supernova remnants, solar flares and active galactic nuclei, and as proposed by \citet{gab69}, their line ratios can be used to diagnose both the electron density and temperature:
\begin{equation}
R(N_e) = \frac{f}{i} = \frac{z}{x+y}
\end{equation}
and
\begin{equation}
G(T_e) = \frac{f + i}{r} = \frac{z + (x + y)}{r}\ .
\end{equation}

\section{Plasma model}
\label{pm}

Following \citet{nes03b}, the stellar coronal plasma is assumed to be optically thin and in collisional equilibrium,
whereby the triplet line emissivities for an ionic species can be determined from the level populations ($N_i$) of an $I_{\rm max}$-level atomic model by solving the statistical equilibrium equation
\begin{equation}\label{pop}
\sum_{k\neq i}N_e N_i q_{ik} + \sum_{k<i} N_i A_{ik} = \sum_{k\neq i}N_e N_k q_{ki} + \sum_{k>i} N_k A_{ki}
\end{equation}
where the total ion density is given by
\begin{equation}
N_{\rm tot}=\sum_{i}^{I_{\rm max}} N_i\ .
\label{Poblaciones}
\end{equation}

In equation~(\ref{pop}), $N_e$ is the electron density, $A_{ki}$ are the spontaneous radiative rates (Einstein $A$-values) and $q_{ik}$ and $q_{ki}$ are, respectively, the excitation and de-excitation generalised rate coefficients \citep{bry05}
\begin{equation}
q_{ik}=\frac{2\sqrt{\pi}\alpha ca_0^2(R/k_B)^{1/2}}{g_i T^{1/2}}
\exp{\left(-\frac{\Delta E_{ik}}{k_BT}\right)}\Upsilon_{ik}
\end{equation}
and
\begin{equation}
q_{ki}=\frac{2\sqrt{\pi}\alpha ca_0^2(R/k_B)^{1/2}}{g_k T^{1/2}}\Upsilon_{ki}\ .
\end{equation}
In these expressions $\Delta E_{ik}=E_k-E_i$ is the energy separation between levels $i$ and $k$ ($i<k)$, $\alpha$ is the fine structure constant, $c$ the speed of light, $a_0$ the Bohr radius, $R$ the Rydberg constant and $k_B$ the Boltzmann constant. $T\equiv 2\overline{E}/3k_B$ is the kinetic temperature which, for a particular normalised electron-energy distribution function $f(E)$, is defined in terms of the mean energy
\begin{equation}
\overline{E} =\int E\,f(E)\mathrm{d}E\ .
\end{equation}
$\Upsilon_{ik}$ and $\Upsilon_{ki}$ are referred to as the effective collision strengths for excitation and de-excitation, respectively, and involve energy averages of the collision strength ($\Omega_{ik}(E)=\Omega_{ki}(E)$) over the electron distribution $f(E)$:
\begin{equation}
\Upsilon_{ik}=\frac{\sqrt{\pi}}{2}\exp{\left(\frac{\Delta E_{ik}}{k_BT}\right)}
\int_0^\infty\Omega_{ik}(\epsilon_i)\left(\frac{\epsilon_i}{k_BT}\right)^{-1/2}f(\epsilon_i)
\mathrm{d}\epsilon_i
\end{equation}
and
\begin{equation}
\Upsilon_{ki}=\frac{\sqrt{\pi}}{2}\int_0^\infty\Omega_{ki}(\epsilon_k)\left(\frac{\epsilon_k}
{k_BT}\right)^{-1/2}f(\epsilon_k)\mathrm{d}\epsilon_k \ .
\end{equation}

\subsection{Maxwell--Boltzmann distribution}
For the MB electron-energy distribution
\begin{equation}
f_{T_e}(E) =\frac{2}{\sqrt{\pi}k_BT_e}\left(\frac{E}{k_BT_e}\right)^{1/2}
\exp{\left(-\frac{E}{k_BT_e}\right)}
\end{equation}
$T_e$ is referred to as the electron temperature, and the effective collision strengths for excitation and de-excitation are equal
\begin{equation}
\Upsilon_{ik}^{\rm MB}(T_e)=\Upsilon_{ki}^{\rm MB}(T_e)=\int_0^\infty \Omega_{ik}(\epsilon_k)\exp{\left(-\frac{\epsilon_k}{k_BT_e}\right)}\,
 \mathrm{d}\left(\frac{\epsilon_k}{k_BT_e}\right)\ ,
\end{equation}
an unmistakable signature of detailed balance.

\subsection{$\kappa$ distribution}

The $\kappa$ distribution takes the form
\begin{equation}
f_{\kappa,E_\kappa}(E) =\frac{2\kappa^{-3/2}}{\sqrt{\pi}E_\kappa}\left(\frac{E}{E_\kappa}\right)^{1/2}
\frac{\Gamma(\kappa+1)}{\Gamma(\kappa-1/2)}
\left(1+\frac{E}{\kappa E_\kappa}\right)^{-(\kappa+1)}
\end{equation}
with a characteristic energy $E_\kappa$ that may be expressed in terms of the kinetic temperature as
\begin{equation}
E_\kappa = k_BT_\kappa(\kappa-3/2)/\kappa\ .
\end{equation}
The $\kappa$ parameter ($3/2<\kappa\leq\infty$) gives a measure of the deviation from the MB distribution, converging to it as $\kappa\rightarrow\infty$. The effective collision strengths for excitation and de-excitation ($i<k$) are no longer equal and are respectively given by
\begin{multline}
\label{kappa_excitation}
\Upsilon_{ik}^\kappa(T_\kappa) = (\kappa-3/2)^{-3/2}
\frac{\Gamma(\kappa+1)}{\Gamma(\kappa -1/2)}\exp{\left(\frac{\Delta E_{ik}}{k_BT_\kappa}\right)}\\
\times\int_0^\infty\Omega_{ik}\left(1+\frac{\epsilon_k+\Delta E_{ik}}
{(\kappa-3/2)k_BT_\kappa}\right)^{-(\kappa+1)}\mathrm{d}\left(\frac{\epsilon_k}{k_BT_\kappa}\right)
\end{multline}
and
\begin{multline}
\label{kappa_deexcitation}
\Upsilon_{ki}^\kappa(T_\kappa) = (\kappa-3/2)^{-3/2}\frac{\Gamma(\kappa+1)}{\Gamma(\kappa -1/2)} \\
\times \int_0^\infty\Omega_{ik}\left(1+\frac{\epsilon_k}
{(\kappa-3/2)k_BT_\kappa}\right)^{-(\kappa+1)}\mathrm{d}\left(\frac{\epsilon_k}{k_BT_\kappa}\right)\ .
\end{multline}

\subsection{Atomic data error propagation}
\label{error_prog}

In previous work on the He-like triplet, the impact of the atomic data errors (i.e. mainly those from the effective collision strengths since the radiative rates are highly accurate) on the line ratios has not been fully studied in either Maxwellian or non-Maxwellian plasmas. Such uncertainties would limit their effectiveness and the temperature and density ranges where they can be applied. We have implemented a numerical method developed by \citet{bau13} to determine their propagation to the values of the level populations, emissivities and line ratios. The associated formalism has been revised and extended, and since some mathematical typos were found, functional expressions for both MB and $\kappa$ distributions are hereby itemised.

The population of the $i$th level in equation~(\ref{pop}) is given by
\begin{equation}
N_i = \frac{\sum_{k\neq i}N_k(N_e q_{ki} + A_{ki})}{N_e\sum_{k\neq i}q_{ik} + \sum_{k < i}A_{ik}}\ ,
\end{equation}
and assuming that the error propagation through the level populations is linear, it may be shown that the level-population relative error for an MB distribution takes the form
\begin{multline}
\label{leverr2}
\left({\delta N_i\over N_i}\right)^2-\sum_{k\ne i} N_k^2{(N_eq_{ki}+A_{ki})^2\over \eta_i^2} \left({\delta N_k\over N_k}\right)^2 \\
={1\over \eta_i^2}\left[N_e^2\sum_{k\ne i}(N_kq_{ki}-N_iq_{ik})^2
\left({\delta \Upsilon_{ki}^{\rm MB}\over \Upsilon_{ki}^{\rm MB}}\right)^2 \right.\\
\left. + \sum_{k>i}(N_kA_{ki})^2\left({\delta A_{ki}\over A_{ki}}\right)^2
+ \left({N_i\over \tau_i}\right)^2 \left({\delta \tau_i\over \tau_i}\right)^2\right]
\end{multline}
where $\tau_i=(\sum_k A_{ik})^{-1}$ is the level lifetime and
\begin{equation}
\eta_i=\sum_{k\ne i}N_k(N_eq_{ki}+A_{ki})\ .
\end{equation}
Similarly, for a $\kappa$ distribution
\begin{multline}
\label{leverr2}
\left({\delta N_i\over N_i}\right)^2 - \sum_{k\ne i} N_k^2{(N_eq_{ki}+A_{ki})^2\over\eta_i^2}\left({\delta N_k\over N_k}\right)^2 \\
= {1\over \eta_i^2}\left[N_e^2\sum_{k\ne i}(N_kq_{ki})^2\left({\delta \Upsilon_{ki}^\kappa\over \Upsilon_{ki}^\kappa}\right)^2 + N_e^2\sum_{k\ne i}(N_iq_{ik})^2 \left({\delta \Upsilon_{ik}^\kappa\over \Upsilon_{ik}^\kappa}\right)^2 \right.\\
\left. + \sum_{k> i}(N_kA_{ki})^2\left({\delta A_{ki}\over A_{ki}}\right)^2
+ \left({N_i\over \tau_i}\right)^2 \left({\delta \tau_i\over \tau_i}\right)^2\right]\ .
\end{multline}

The relative error of the line emissivity $j_{ki}=N_kA_{ki}\Delta E_{ik}$, where $k>i$, is then given by
\begin{multline}
\left({\delta j_{ki}\over j_{ki}}\right)^2 =
\left({\delta N_k\over N_k}\right)^2
+\left({N_k\over \eta_k\tau_k}\right)^2\left({\delta\tau_k \over \tau_k}\right)^2 \\
+\left(1-{N_k\over \eta_k}A_{ki}\right)^2
\left({\delta A_{ki}\over A_{ki}}\right)^2\ ;
\end{multline}
furthermore, for the line ratio ($k>i$ and $h>g$)
\begin{equation}
D = \frac{j_{ki}}{j_{hg}} = \left(\frac{N_k}{N_h}\right) \left(\frac{A_{ki}}{A_{hg}}\right) \left(\frac{\Delta E_{ik}}{\Delta E_{gh}}\right)
\end{equation}
the relative error can be expressed as
\begin{multline}
\label{lru}
\left(\frac{\delta D}{D}\right)^2 =  \left[1-D\left({A_{hg}\Delta E_{gh}\over
A_{ki}\Delta E_{ik}}{\partial N_h\over \partial N_k} +
{A_{hg}\Delta E_{gh}\over \Delta E_{ik} N_k}
{\partial N_h\over \partial A_{ki}}\right)\right]^2\left(
{\delta j_{ki}\over j_{ki}}\right)^2
\\ +
\left[1-\frac{1}{D}\left({A_{ki}\Delta E_{ik}\over
A_{hg}\Delta E_{gh}}{\partial N_k\over \partial N_h} +
{A_{ki}\Delta E_{ik}\over \Delta E_{gh} N_h}
{\partial N_k\over \partial A_{hg}}\right)\right]^2\left(
{\delta j_{hg}\over j_{hg}}\right)^2\ .
\end{multline}
In accord with \citet{bau13}, we find that $\partial N_k/\partial A_{hg}=N_kN_h/\eta_k$ for $g=k$; $\partial N_k/\partial A_{hg}=-N_k^2/\eta_k$ for $h=k$ and $\partial N_k/\partial A_{hg}=0$ otherwise.
Also, expression~(\ref{lru}) can be extended to a line ratio
\begin{equation}
D = {\sum_{\{ki\}} j_{ki}\over \sum_{\{hg\}} j_{hg}}
\end{equation}
with multiple components where $\{ki\}$ now denotes a summation over line pairs:
\begin{multline}
\left(\frac{\delta D}{D}\right)^2 = \sum_{\{ki\}} \left(
{\sum_{\{ki\}'} (\partial j_{\{ki\}'}/\partial j_{\{ki\}})\over
\sum_{\{ki\}} j_{\{ki\}}}
- {\sum_{\{hg\}} (\partial j_{\{hg\}}/\partial j_{\{ki\}})\over
\sum_{\{hg\}} j_{\{hg\}}}
\right)^2 (\delta j_{\{ki\}})^2
\\ +
\sum_{\{hg\}} \left(
{\sum_{\{ki\}} (\partial j_{\{ki\}}/\partial j_{\{hg\}})\over
\sum_{\{ki\}} j_{\{ki\}}}
- {\sum_{\{hg\}'} (\partial j_{\{hg\}'}/\partial j_{\{hg\}})\over
\sum_{\{hg\}} j_{\{hg\}}} \right)^2 (\delta j_{\{hg\}})^2\ .
\label{Error-variosL}
\end{multline}

\section{Atomic data}
\label{atomic_data}

One of the main objectives of the present work is to appraise the accuracy of the available atomic data for members of the He isoelectronic sequence with atomic number $Z\leq 14$, namely the $A$-values and effective collision strengths that go into equation~(\ref{pop}). Regarding the radiative data for an atomic model that includes the 49 fine-structure levels with principal quantum number $n\leq 5$ in \ion{C}{v}, \ion{N}{vi}, \ion{O}{vi}, \ion{Ne}{ix}, \ion{Mg}{xi} and \ion{Si}{xiii}, we adopt the $A$-values computed with a relativistic configuration-interaction method by \citet{sav03}; they take into account electric dipole (E1), electric quadrupole (E2), magnetic dipole (M1) and magnetic quadrupole (M2) transitions. On the other hand, the two-photon decay rates of the $2\,^1{\rm S}_0$ metastable level are by \citet{dra86} who used a relativistic method based on correlated variational wave functions of the Hylleraas type. These transition-probability data sets have been shown to be accurate to better than 3\% by \citet{men14b}. Therefore, the key benchmark is the accuracy of the effective collision strengths for these ionic species reported by \citet{zha87}, \citet{del02}, \citet{che06}, \citet{agg08, agg10, agg12} and  \citet{agg09, agg11}.

\begin{figure}
  \begin{minipage}[b]{0.5\textwidth}
  \centering
\includegraphics[width=0.85\columnwidth]{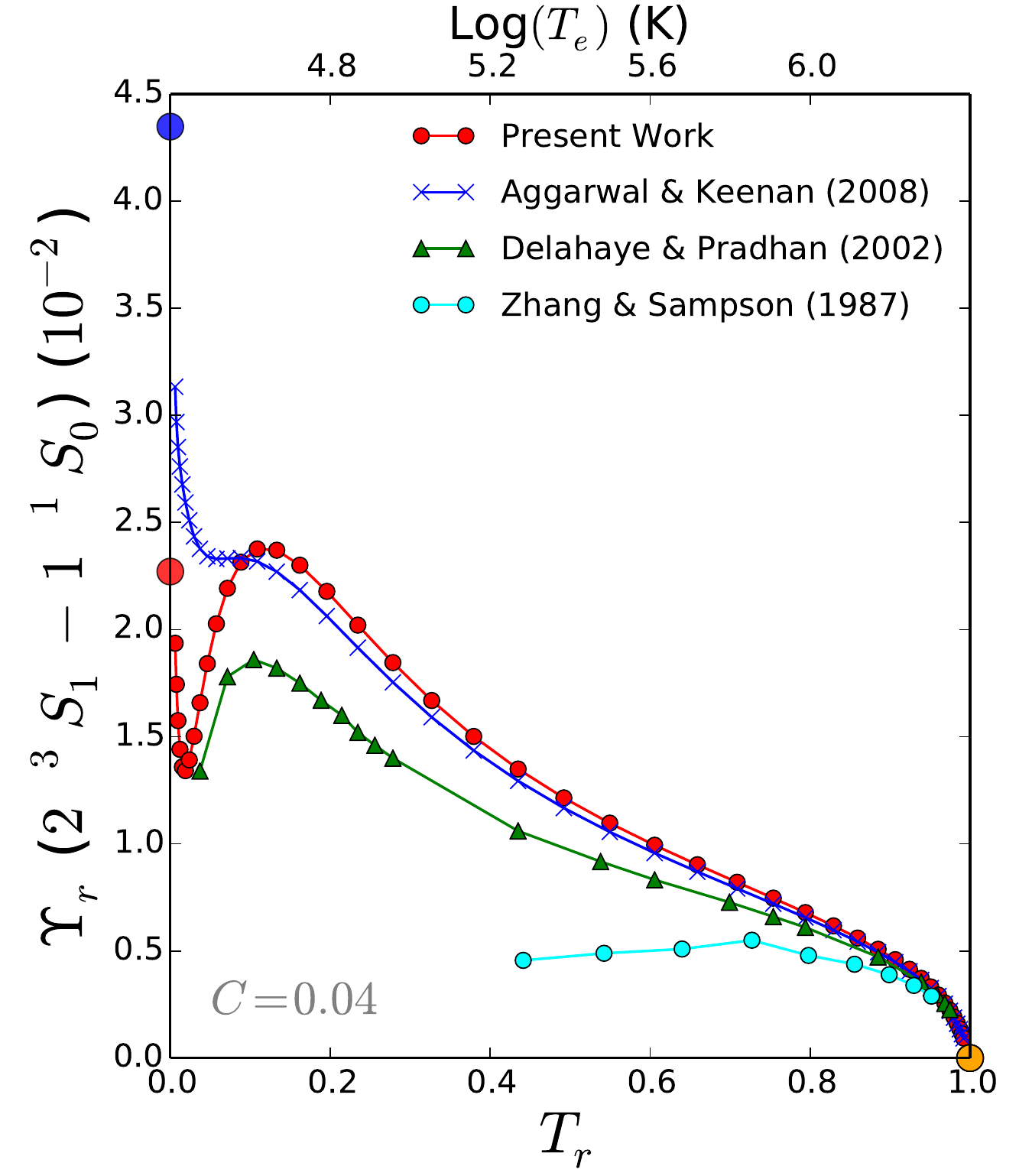}
  \end{minipage}
  \begin{minipage}[b]{0.5\textwidth}
  \centering
  \includegraphics[width=0.85\columnwidth]{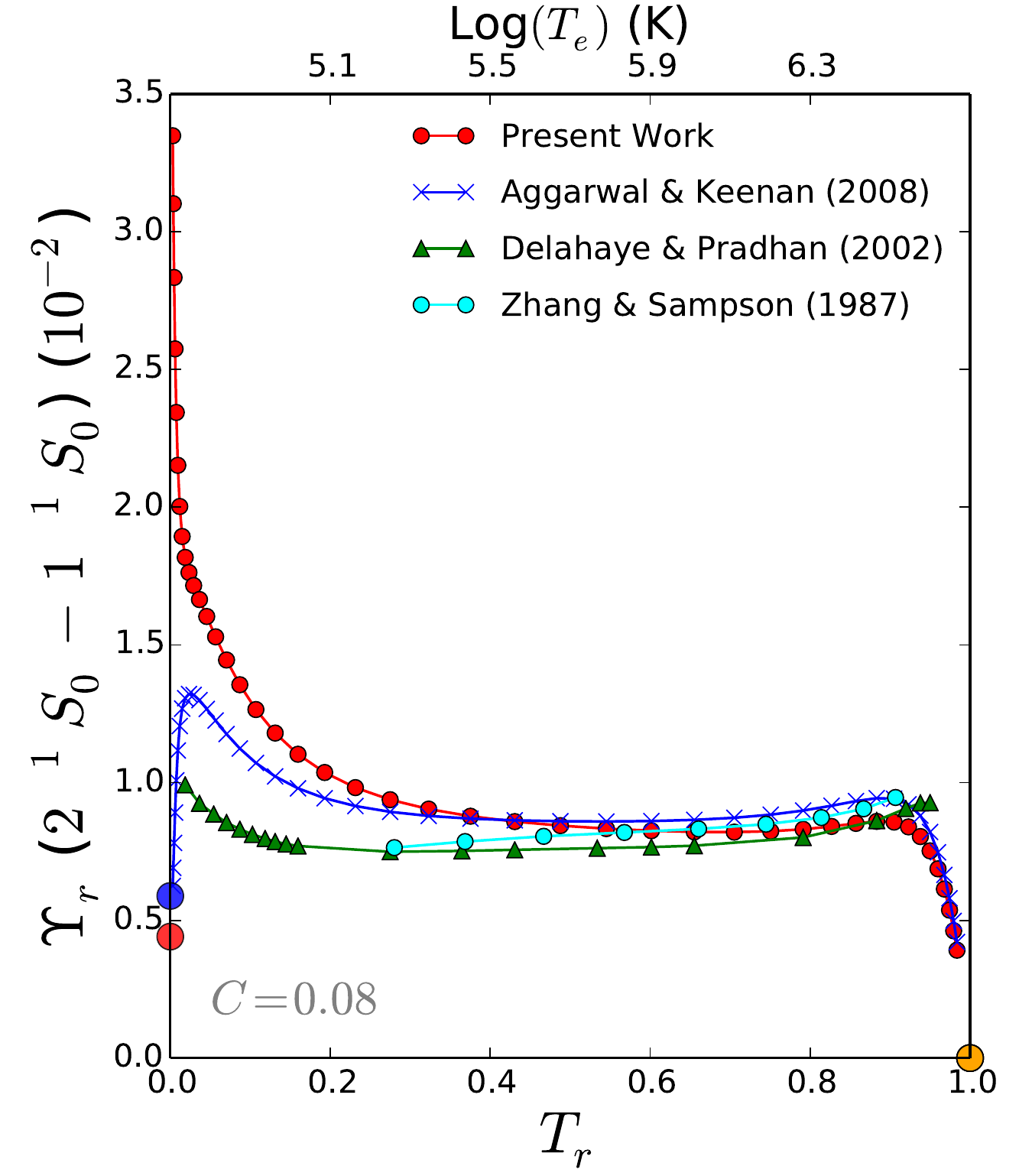}
  \end{minipage}
  \caption{Reduced effective collision strengths for the transitions ${\rm 2\,^3S_1 - 1\,^1S_0}$  (upper panel) and ${\rm 2\,^1S_0 - 1\,^1S_0}$  (lower panel) in \ion{O}{vii}. Red circles: present results. Blue crosses: \citet{agg08}. Green triangles: \citet{del02}. Cyan circles: \citet{zha87}. $\Upsilon_r(0)$ are obtained from the energy tabulation of the corresponding collision strength as $\Upsilon_r(0)=\Omega(0)$, and $\Upsilon_r(1)$ (yellow circles) are estimated with  {\sc autostructure}. The $C$ fit parameter for each plot is indicated. \label{o7_ups1}}
\end{figure}

\begin{figure}
  \centering
  \includegraphics[width=0.85\columnwidth]{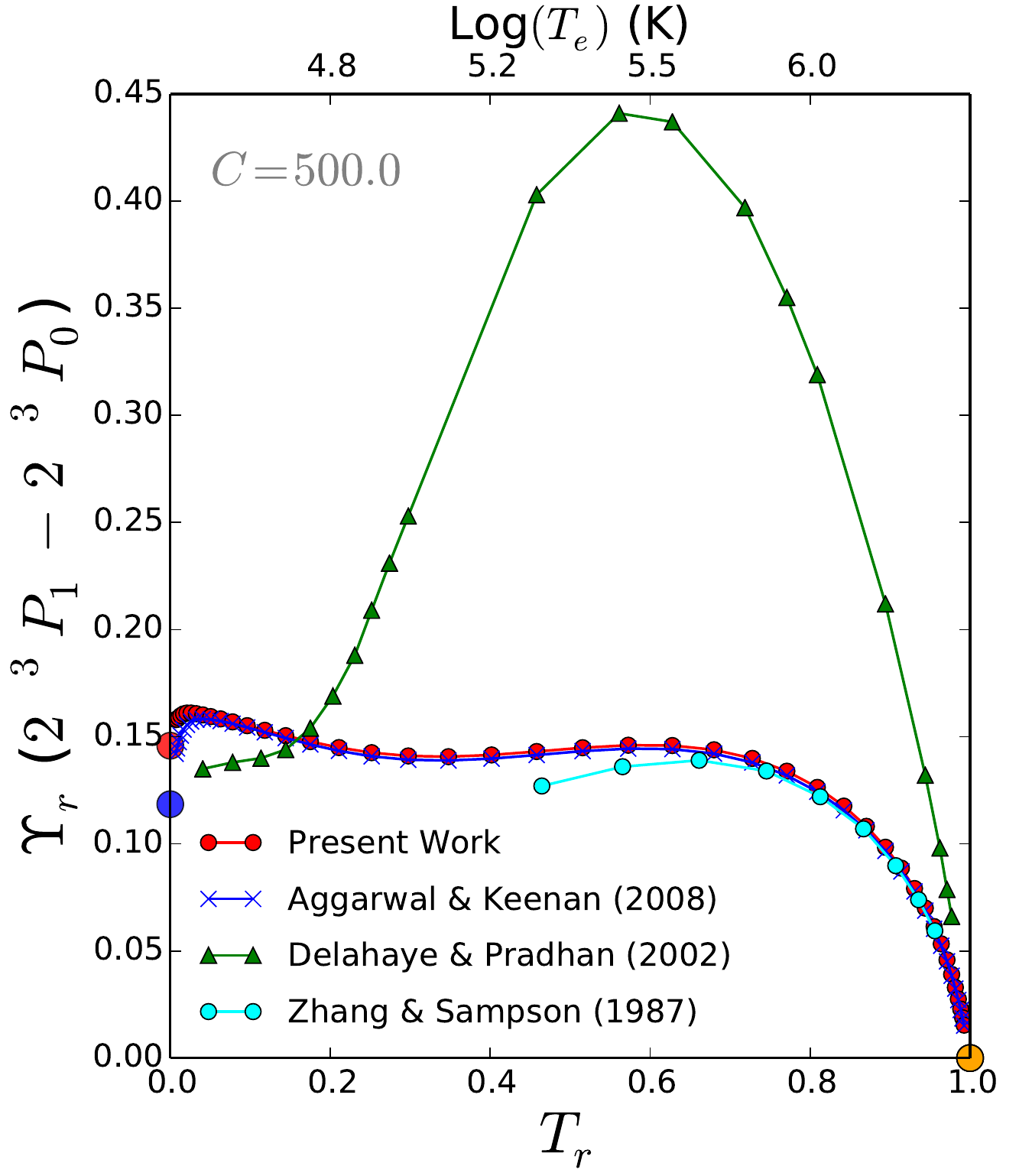}
  \caption{Reduced effective collision strength for the ${\rm 2\,^3P^o_1 - 2\,^3P^o_0}$ transition in \ion{O}{vii}. Red circles: present results. Blue crosses: \citet{agg08}. Green triangles: \citet{del02}. Cyan circles: \citet{zha87}. $\Upsilon_r(0)$ are obtained from the energy tabulation of the corresponding collision strength as $\Upsilon_r(0)=\Omega(0)$, and $\Upsilon_r(1)$ (yellow circles) is estimated with {\sc autostructure}. A $C=500.0$ fit parameter was used. \label{o7_ups2}}
\end{figure}

\begin{figure*}
  \centering
  \begin{minipage}[b]{0.32\textwidth}
  \includegraphics[width=\columnwidth]{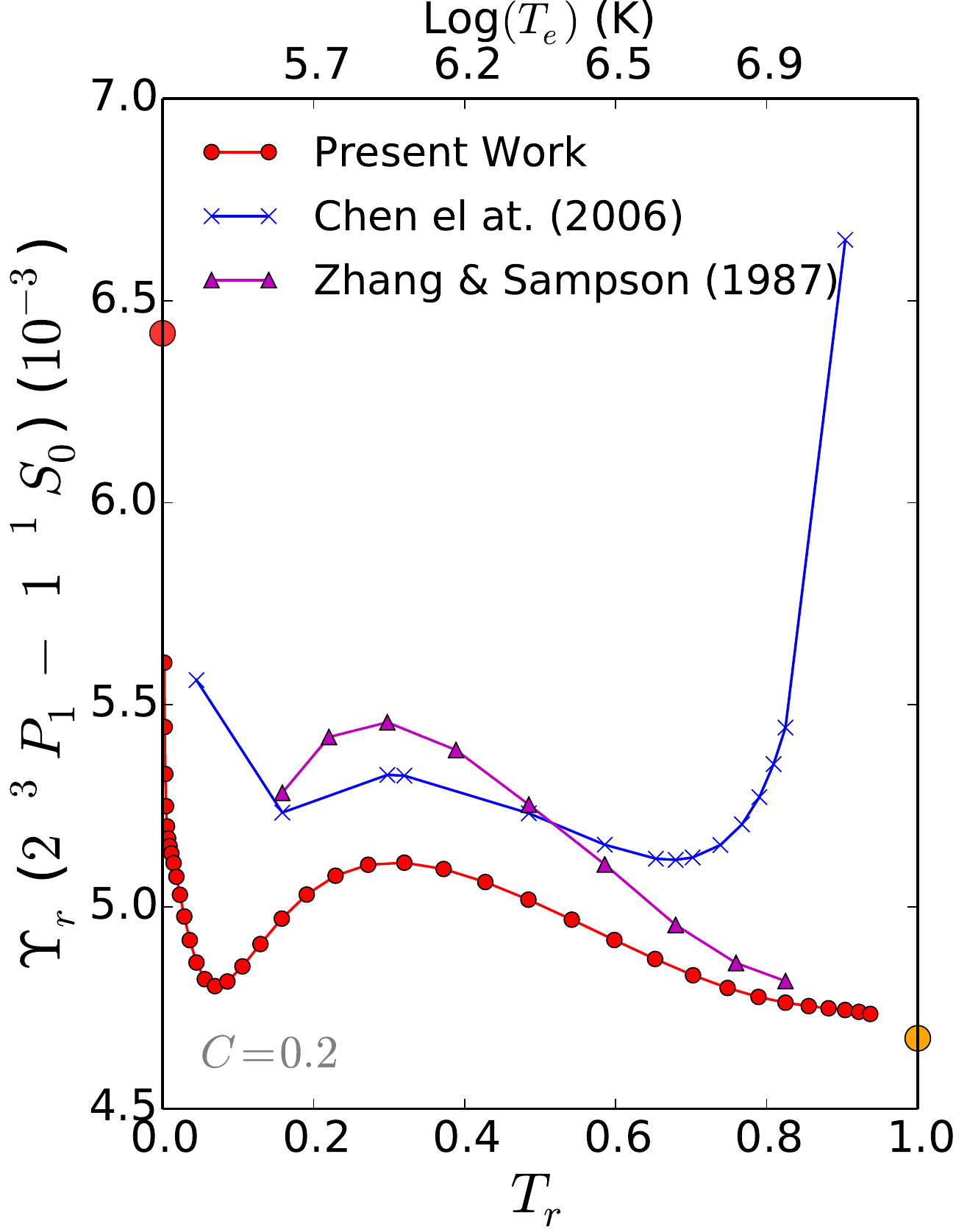}
  \end{minipage}
  \hfill
  \begin{minipage}[b]{0.32\textwidth}
  \includegraphics[width=\columnwidth]{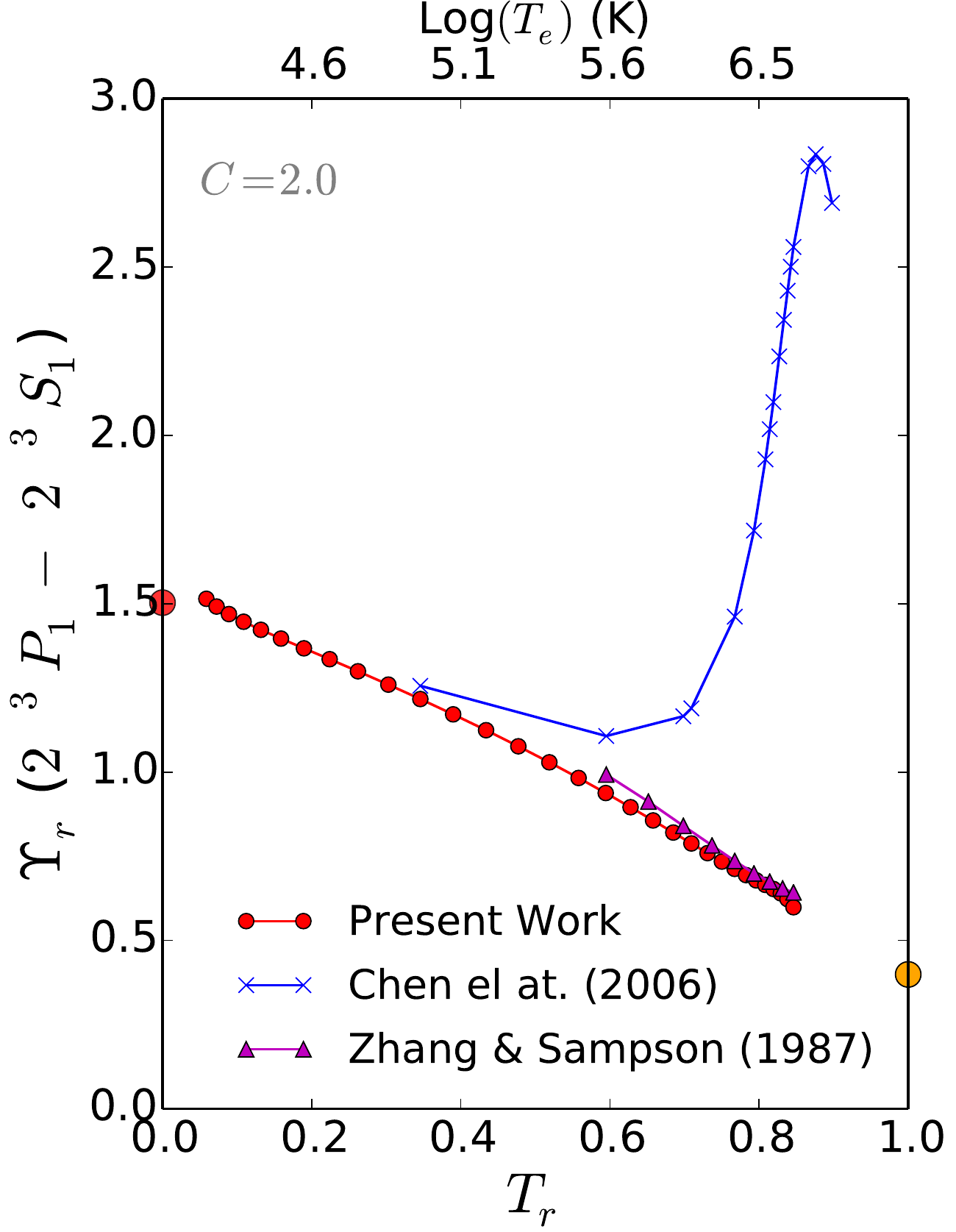}
  \end{minipage}
  \hfill
  \begin{minipage}[b]{0.32\textwidth}
  \includegraphics[width=\columnwidth]{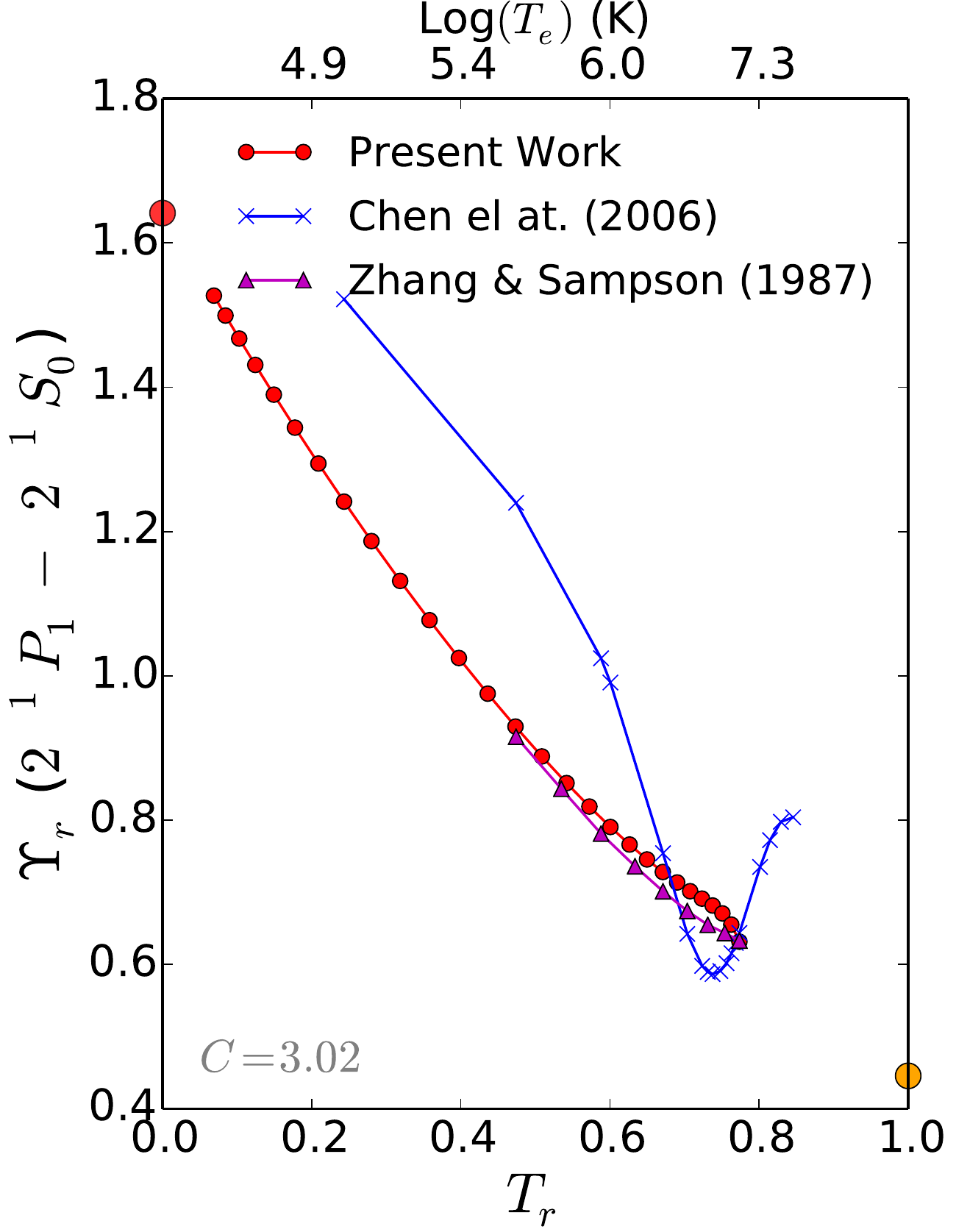}
  \end{minipage}
  \caption{Reduced effective collision strengths for the transitions ${\rm 2\,^3P^o_1 - 1\,^1S_0}$ (left panel), ${\rm 2\,^3P^o_1 - 2\,^3S_1}$ (middle panel) and ${\rm 2\,^1P^o_1 - 2\,^1S_0}$ (right panel) in \ion{Ne}{ix}. Red circles: present results. Blue crosses: \citet{che06}. Purple triangles: \citet{zha87}. $\Upsilon_r(0)$ are obtained from the energy tabulations of the collision strength as $\Upsilon_r(0)=\Omega(0)$, and $\Upsilon_r(1)$ (yellow circles) are estimated with  {\sc autostructure}. The $C$ fit parameter for each plot is indicated. \label{ne9_ups}}
\end{figure*}

 The evaluation of the effective collision strengths is carried out by means of the formalism developed by \citet{bur92} wherein, according to the transition type (allowed, intercombination or forbidden), the effective collision strength and electron temperature are scaled such that the temperature range $0\leq T_e\leq \infty$ is reduced to $0\leq T_r\leq 1$. Important fit parameters are: $\Upsilon_r(0)=\Omega(0)$ determined from the collision strength tabulation; $\Upsilon_r(1)$ have been obtained with the {\sc autostructure} atomic structure code \citep{eis74, bad11} and $C$ which is adjusted to distribute the points evenly along the reduced temperature range. This subtle scaling procedure brings out the effective-collision-strength behaviour in both the low- and high-temperature regimes.

We choose \ion{O}{vii} as a case study since this system has received the most computational attention: effective collision strengths for transitions among $n\leq 2$ levels obtained with the Coulomb--Born exchange method \citep{zha87}; those for  $n\leq 4$ transitions calculated with the Breit--Pauli $R$-matrix method \citep[{\sc bprm},][]{del02} and, among $n\leq 5$ levels, those computed with the Dirac Atomic $R$-matrix method \citep[{\sc darc},][]{agg08}. For the latter we have had access to the collision-strength energy tabulations that enable more thorough comparisons, although we only had available the {\sc bprm} effective collision strengths from \citet{del02} for transitions involving levels with $n\leq 2$. It is worth emphasising that all these calculations take into account the important resonance effects. Furthermore, the present detailed comparison between {\sc bprm} and {\sc darc} effective collision strengths for two-electron targets leads to a timely milestone in the current polemic around the outstanding inconsistencies that have been found in some of the results by these two methods for Be-like ions \citep{agg15a, agg15b, fer15}.

Due to noticeable differences in the effective collision strengths by \citet{del02} and \citet{agg08} for some transitions among levels within $n=2$ in \ion{O}{vii}, we were encouraged to calculate collision strengths for this system with {\sc bprm} in order to resolve this disaccord. The {\sc bprm} scattering method is fully described by \citet{ber95}. We developed a 49-level ($n\leq 5$) target with {\sc autostructure}, and included experimental target level energies in {\sc bprm} at Hamiltonian diagonalisation. Contributions from all partial waves with total angular momentum $J\leq 18.5$ have been taken into account plus complements from top-up procedures for higher $J$; collision strengths in the resonance region were computed with a fine energy-mesh step of $\Delta E=3.6\times 10^{-4}$~Ryd.

The general level of agreement of our new effective collision strengths for \ion{O}{vii} with those computed by \citet{agg08} for transitions between $n\leq 3$ levels is better than 5\% and within 10\% with those by \citet{zha87} for $n\leq 2$. The larger discrepancies are found at very low temperatures (irrelevant to the He-like triplet diagnostics) in forbidden transitions with small effective collision strengths ($\Upsilon\lesssim 10^{-2}$) where the resonance contribution is dominant. As shown in Fig.~\ref{o7_ups1}, this is the case of the transitions ${\rm 2\,^3S_1 - 1\,^1S_0}$  and ${\rm 2\,^1S_0 - 1\,^1S_0}$ at $T_r < 0.2$, particularly as $T_r\rightarrow 0$. As specified by \citet{bur92}, the reduced effective collision strength strength at $T_r=0$ is obtained from the energy tabulation of the collision strength, $\Upsilon_r(0)=\Omega(0)$. The huge differences at  very low reduced temperatures (as large as a factor of 2 for the ${\rm 2\,^3S_1 - 1\,^1S_0}$ transition) are caused by small shifts in the resonance energy positions ($\Delta\epsilon_j\approx 0.015$~Ryd) that lead to incongruent resonance structures. In this case they are believed to arise from using experimental instead of theoretical threshold energies in our {\sc bprm} Hamiltonian.

The agreement with the data by \citet{del02} for \ion{O}{vii} is also comparable: $\sim 8\%$ at the temperatures of interest (see Fig.~\ref{o7_ups1}), but as shown in Fig.~\ref{o7_ups2}, large enhancements (a factor of 3) are found in their effective collision strengths for the transitions ${\rm 2\,^3P^o_J - 2\,^3P^o_{J^{'}}}$ at the higher temperatures. In contrast to \citet{agg08} and the present calculation, both \citet{zha87} and \citet{del02} included radiation damping, but taking into consideration our good agreement with \citet{zha87} for these transitions, such large increments are not expected to be due to this effect that is usually small in low-$Z$ ionic species.

Effective collision strengths by \citet{agg08} for transitions between levels with $4\leq n\leq 5$ in \ion{O}{vii} are on average within 12\% of the present results, except for allowed and intercombination transitions between quasi-degenerate levels, where in many cases the theoretical level-energy order does not agree with the NIST spectroscopic tables. Such levels are usually treated as degenerate in {\sc bprm} calculations. The discrepancies are found to be large: a factor of $\sim 4$ for $4\,^1{\rm F}^{\rm o}_3-4\,^1{\rm D}_2$; a factor of $\sim 2$  for $4\,^3{\rm D}_3-4\,^3{\rm F}^{\rm o}_4$; a factor of $\sim 1.5$ for $4\,^1{\rm D}_2-4\,^3{\rm F}^{\rm o}_3$; a factor of 22 for $5\,^3{\rm G}_3-5\,^3{\rm P}^{\rm o}_2$ and factors as large as 9 for $5\,^3G_J-5\,^3F^{\rm o}_{J'}$. In this context, our adopted partial wave expansion, which works adequately for most slowly converging transitions (i.e. E1 transitions), is not extensive enough as to allow us to explain with confidence the cause of these differences.

The average agreement between the effective collision strengths by \citet{zha87} for transitions between $n\leq 2$ levels with \citet{agg12} for \ion{Mg}{xii} and \citet{agg10} for \ion{Si}{xiii} is within 2\%, except again for the ${\rm 2\,^3{\rm S}_1 - 1\,^1{\rm S}_0}$ forbidden transition where differences as large as a factor of 2 are observed at low temperatures. On the other hand, remarkable discrepancies are found between \citet{zha87} and \citet{che06} for several $n\leq 2$ transitions in \ion{Ne}{ix} at high temperatures, specially for allowed and intercombination transitions (see Fig.~\ref{ne9_ups}). Due to this disaccord, we have computed effective collision strengths for this ion with {\sc bprm}, some of which are also displayed in Fig.~\ref{ne9_ups}. Our calculation again introduces spectroscopic target energies at Hamiltonian diagonalisation; takes into account partial wave contributions with total angular momentum $J\leq 36.5$, topping up for higher $J$; and implements an energy-mesh step of $\Delta E=6.4\times 10^{-4}$~Ryd. The average level of agreement between the present effective collision strengths with those by \citet{zha87} for $n\leq 2$ transitions in \ion{Ne}{ix} is within 2\% if the ${\rm 2\,^3{\rm S}_1 - 1\,^1{\rm S}_0}$ forbidden transition is excluded. In general, we find discrepancies with the effective collision strengths by \citet{che06} for many $n\leq 4$ transitions, which therefore points out at unreliable data.

\citet{zha87} did not consider \ion{C}{v} and \ion{N}{vi}; thus the only available effective collision strengths for these ions are by \citet{agg11} and \citet{agg09}, respectively. In order to evaluate their accuracy, we have computed effective collision strengths for these ions with the {\sc bprm} method using very similar models to that for \ion{O}{vii}. For transitions within $n\leq 3$, the agreement was find to be within 10\% except for allowed transitions between quasi-degenerate levels, e.g.  $3\,^3{\rm P}^{\rm o}_J - 3\,^3{\rm S}_1$, for which the differences are found to be as large as a factor of 2 at high temperatures.

\section{Diagnostics for MB distribution}
\label{MB}

Treating \ion{O}{vii} again as a case study, we have derived the $R(N_e)$ and $G(T_e)$ emissivity ratios by solving equation~(\ref{pop}) for different atomic models ($2\leq n\leq 5$), including the $A$-values of \citet{sav03} and \citet{dra86} and the effective collision strengths by \citet{agg08}. The $R(N_e)$ and $G(T_e)$ uncertainty intervals have been estimated from the atomic data errors using the formalism developed in Section~\ref{error_prog}. The $A$-value errors have been globally estimated to be within 3\% \citep{men14b}, while a temperature-dependent error margin of the effective collision strength for each transition has been obtained from the ratios of \citet{agg08} and the present values.

\begin{figure}
  \centering
  \includegraphics[width=\columnwidth]{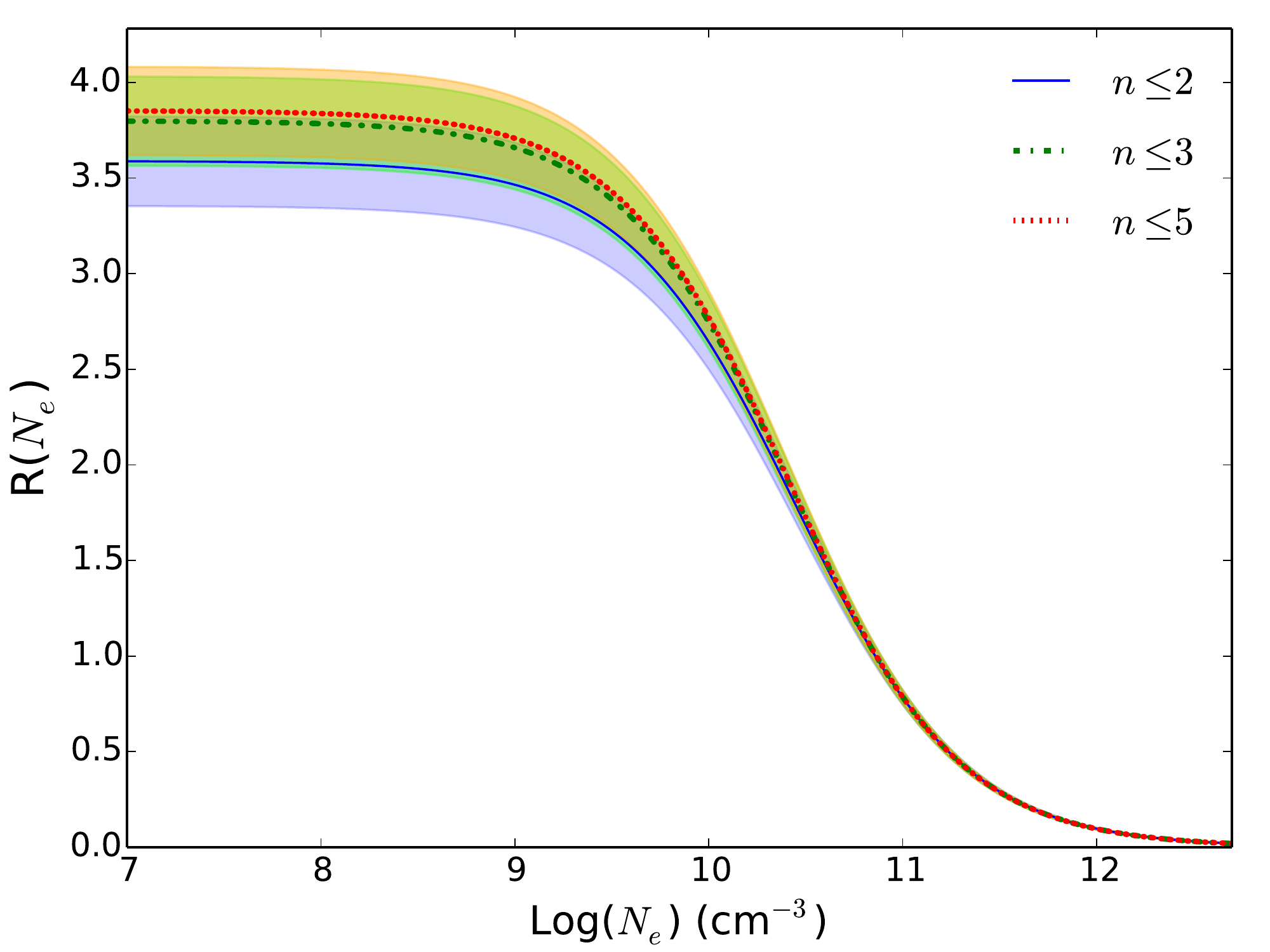}
  \caption{$R(N_e)$ density sensitive line ratio for different atomic models of \ion{O}{vii} assuming an MB electron-energy distribution at $T_e=1.0$~MK. Blue continuous curve: 7-level model ($n\leq 2$). Green dash-dotted curve: 17-level model ($n\leq 3$). Red dotted curve: 49-level model ($n\leq 5$). \label{R-ratio}}
\end{figure}

\begin{figure}
  \centering
  \includegraphics[width=\columnwidth]{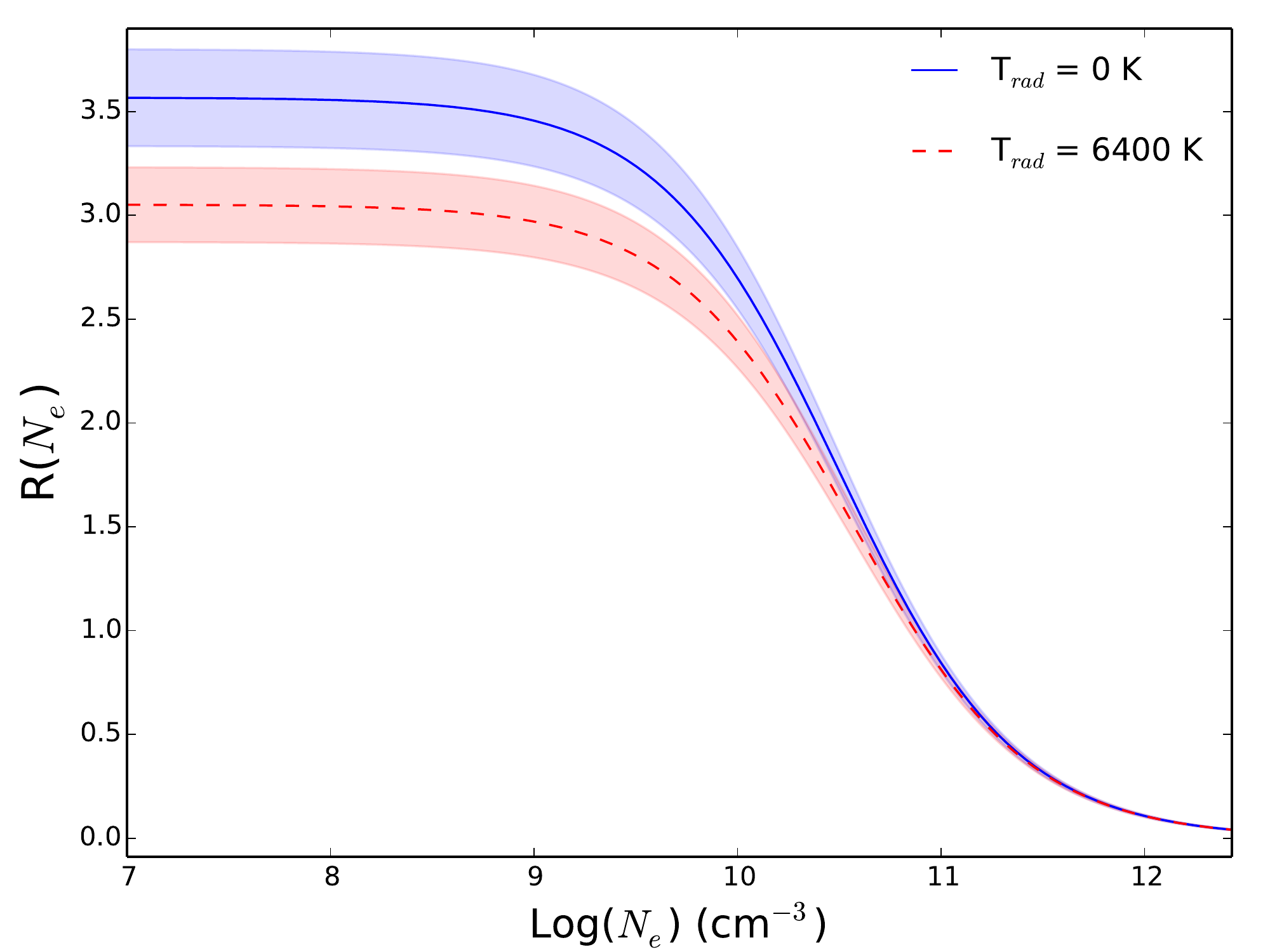}
  \caption{$R(N_e)$ density sensitive line ratio in \ion{O}{vii}  showing the decrease of the low-density limit caused by taking into account a level-populating radiation field in equation~(\ref{pop}). Blue continuous curve: no radiation field. Dashed red curve: including a radiation field with $T_{\rm rad}=6400$~K. An MB electron-energy distribution at $T_e=1.0$~MK is assumed. \label{R-rad}}
\end{figure}

The $R(N_e)$ density sensitive line ratio is plotted in Fig.~\ref{R-ratio} for different \ion{O}{vii} atomic models in the electron-density range $7\leq \log(N_e)\leq 13$~cm$^{-3}$. It may be seen that only the low-density region is sensitive to the atomic model, and convergence is well reached in a model that includes levels with $n\leq 3$. Also, line-ratio error bands of $\sim 6\%$ are only conspicuous in the low-density regime, which would compromise the $R(N_e)$ diagnostic capability of discerning the effects of a radiation field as an additional level populating mechanism in equation~(\ref{pop}). As discussed by \citet{nes01, nes02a} and \citet{por02}, the inclusion of the radiation field lowers the low-density limit, and if the line-ratio uncertainty margin is taken into account, $R(N_e)$ would not be able to differentiate its effect in \ion{O}{vii} for radiation temperatures $T_{\rm rad}\lesssim 6400$~K (see Fig.~\ref{R-rad}).

\begin{figure}
  \centering
  \includegraphics[width=\columnwidth]{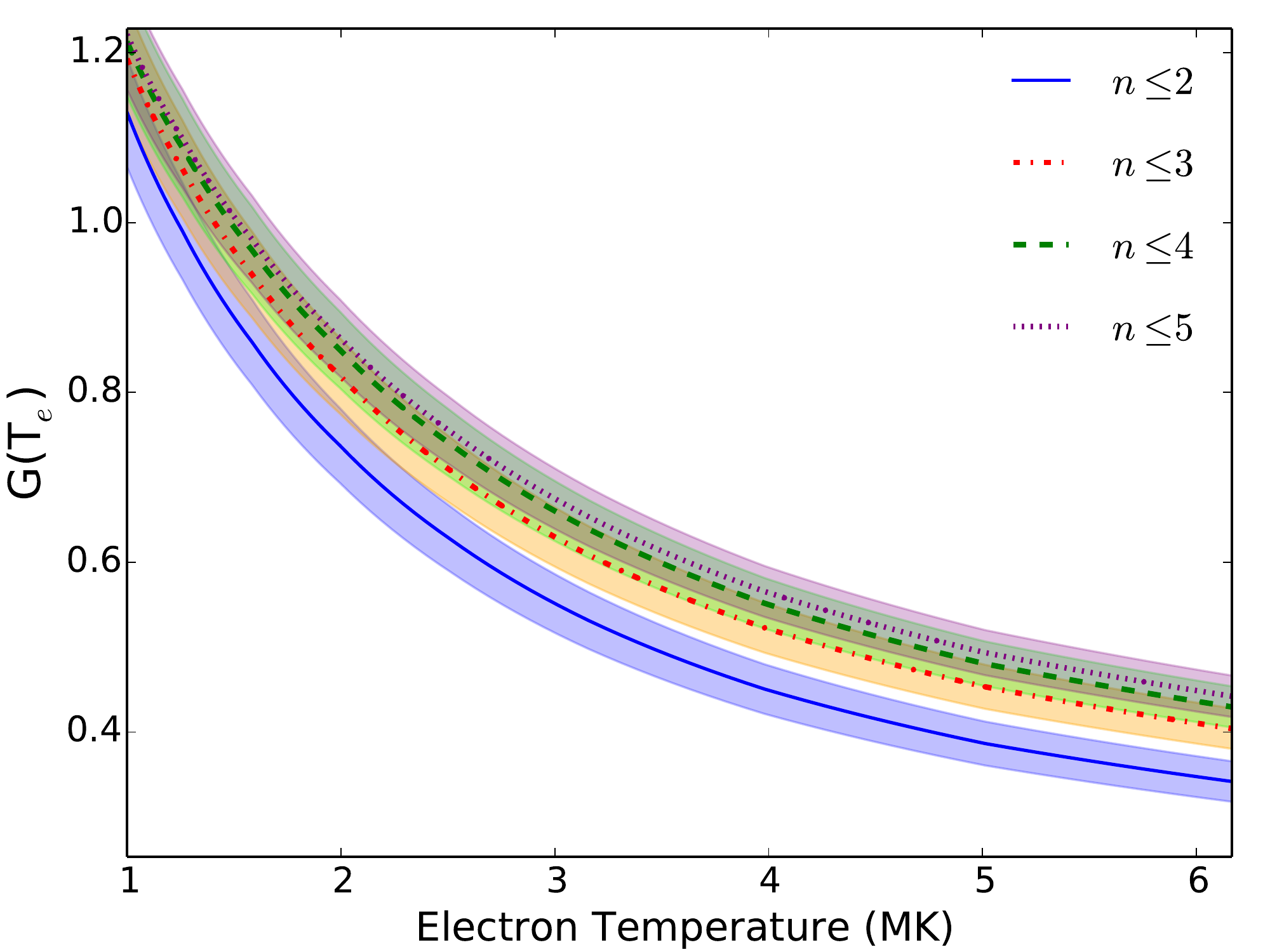}
  \caption{$G(T_e)$ temperature sensitive line ratio in \ion{O}{vii} for different atomic models assuming an MB electron energy distribution with $N_e=10^{10}$~cm$^{-3}$. Blue continuous curve: 7-level model ($n\leq 2$). Red dash-dotted curve: 17-level model ($n\leq 3$). Dashed green curve:  35-level model ($n\leq 4$). Purple dotted curve:  49-level model ($n\leq 5$). \label{G-ratio}}
\end{figure}

Similarly, the $G(T_e)$ temperature sensitive ratio for this system is plotted in Fig.~\ref{G-ratio} in the $1.0\leq T_e\leq 6.0$~MK electron-temperature range. If the error margins are taken into account, this line ratio is insensitive to the atomic model beyond $n=3$, and would be practically ineffective as a temperature diagnostic for $T_e> 5$~MK. The convergence of the $R$ and $G$ line ratios with an atomic model based on $n\leq 3$ levels is also found for other members of the He isoelectronic sequence with $Z\leq 14$.

\begin{figure*}
  \centering
  \includegraphics[width=15cm]{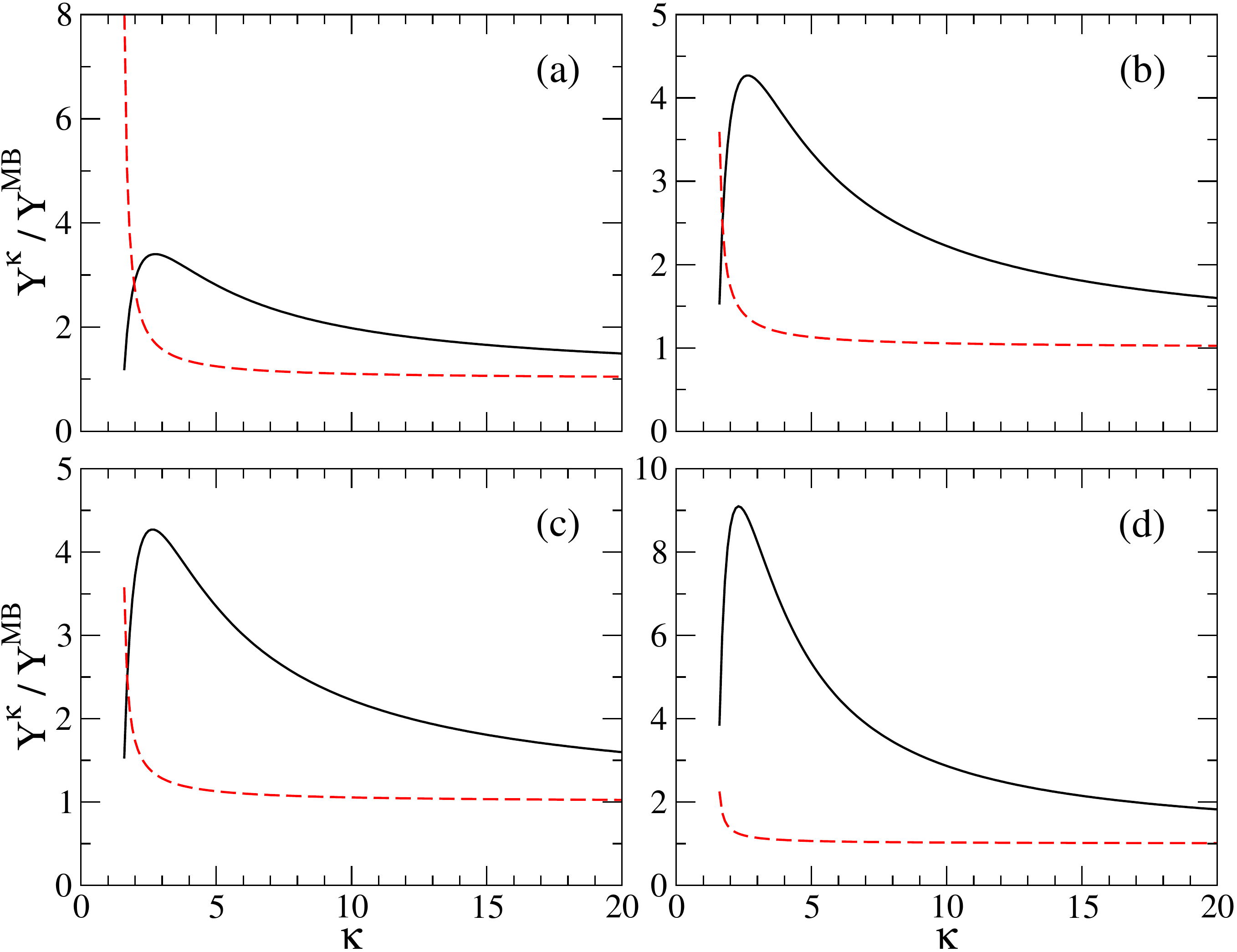}
  \caption{Effective-collision-strength ratio $\Upsilon^\kappa/\Upsilon^{\rm MB}$ at $T=1$~MK as a function of $\kappa$. Black continuous line: excitation.  Red dashed line: de-excitation. (a) $2\,^3{\rm S}_1-1\,^1{\rm S}_0$ magnetic dipole transition. (b) $2\,^3{\rm P^o}_1-1\,^1{\rm S}_0$ intercombination transition. (c) $2\,^3{\rm P^o}_2-1\,^1{\rm S}_0$ magnetic quadrupole transition. (d) $2\,^1{\rm P^o}_1-1\,^1{\rm S}_0$ electric dipole allowed transition. \label{ups_kappa}}
\end{figure*}

\begin{figure}[!h]
  \centering
  \includegraphics[width=\columnwidth]{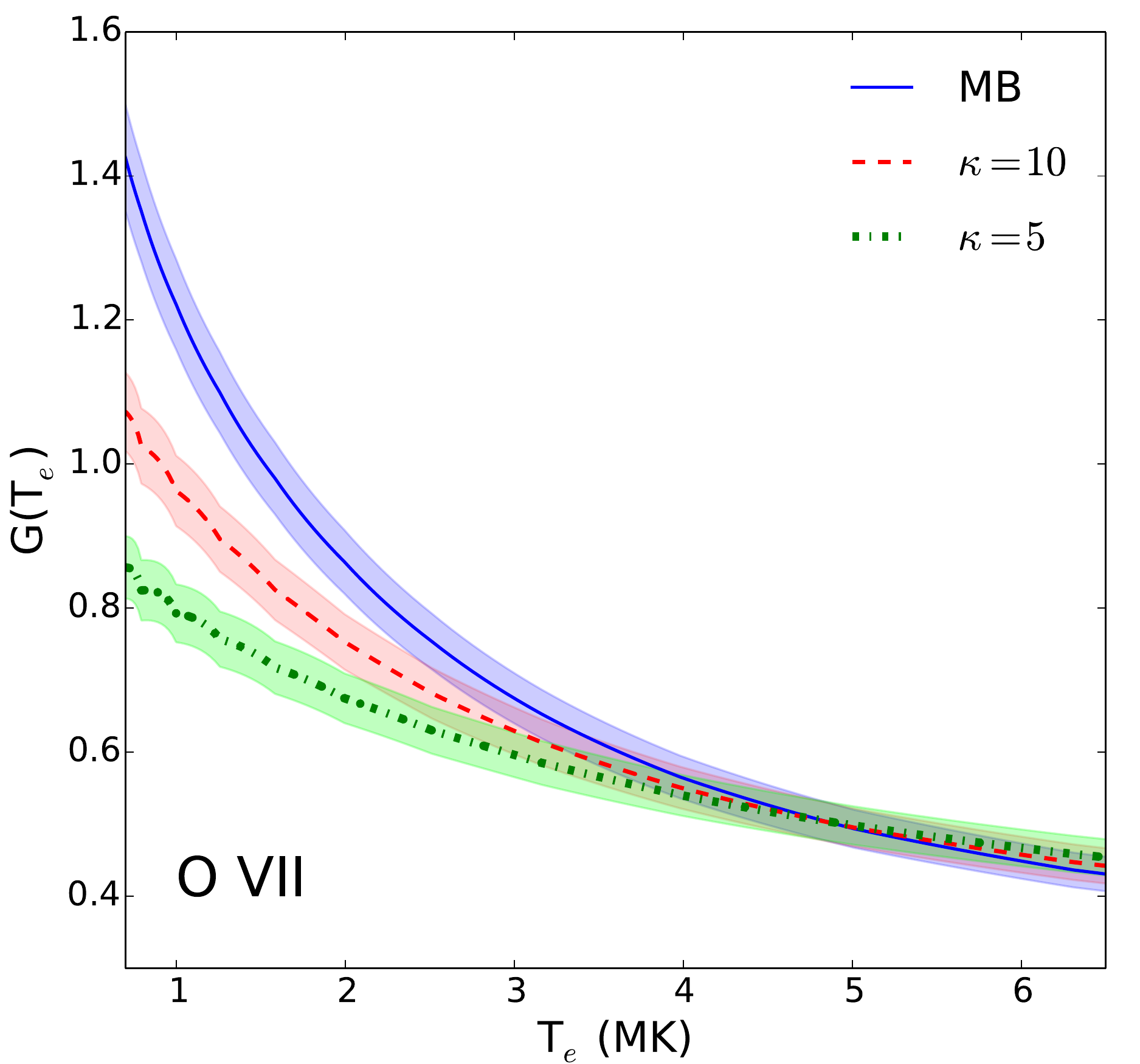}
  \caption{$G(T_e)$ temperature sensitive line ratio in \ion{O}{vii} computed with a 49-level model ($n\leq 5$) at $N_e=10^{10}$~cm$^{-3}$. Blue continuous curve: MB distribution is assumed. Red dashed curve: $\kappa=10$ distribution.  Green dash-dotted curve: $\kappa=5$ distribution. \label{G-kappa}}
\end{figure}

\section{Diagnostics for $\kappa$ distribution}
\label{kappa}

In order to determine the $R(N_e)$ and $G(T)$ line ratios assuming a $\kappa$ distribution, $\kappa$-averaged effective collision strengths have been calculated for excitation and de-excitation with equations~(\ref{kappa_excitation})--(\ref{kappa_deexcitation}). These integrals have been estimated from the energy tabulations of the collision strengths of \citet{agg08, agg10, agg12} and  \citet{agg09, agg11} for \ion{C}{v}, \ion{N}{vi}, \ion{O}{viii}, \ion{Mg}{xi} and \ion{Si}{xiii}, and from the present calculation for \ion{Ne}{ix}. In Fig.~\ref{ups_kappa}, the ratio of the $\kappa$-averaged effective collision strength to the MB value, $\Upsilon^\kappa/\Upsilon^{\rm MB}$, at $T=1$~MK is plotted for the four transitions of the He-like triplet in \ion{O}{vii}. For de-excitation $\Upsilon^\kappa/\Upsilon^{\rm MB}\approx 1$ except for $\kappa < 2$ when it abruptly increases by several factors. For excitation, this ratio steadily increases for $\kappa < 10$ reaching a maximum at $\kappa \approx 2.5$ whose value depends on the transition energy $\Delta E_{ik}$; i.e. the largest ratio (factor of $\sim 9$) corresponds to the dipole allowed resonance transition with the shortest wavelength.

It is found that the adoption of a $\kappa$ distribution hardly changes the $R(N_e)$ line ratio; only small decrements ($\sim3\%$) are detected in the low-density limit that lie within the line-ratio error margin. On the other hand, as shown in Fig.~\ref{G-kappa}, the $\kappa$ distribution significantly reduces the $G(T)$ at the lower temperatures in \ion{O}{vii} from the MB value (60\% for $\kappa=5$ at 1~MK). A similar behaviour is found in all of the He-like ions ($6\leq Z\leq 14$) considered in the present work; and there is a $Z$-increasing maximum temperature along the isoelectronic sequence (e.g. 5~MK in \ion{O}{vii}) for which the $G(T)$ line ratio of the $\kappa$ distribution cannot be differentiated from the MB, particularly when the error margins are taken into account. It may also be seen that at low temperatures the $\kappa$ line ratio tends to oscillate; this occurs when solving equation~(\ref{pop}) with very large $\kappa$-averaged effective collision strengths.

\section{Observational benchmark}
\label{obs}

In order to substantiate our assessment of the accuracy of the atomic data that are used to model the He-like triplet emissivities and its impact on the $R(N_e)$ and $G(T)$ line ratios in both Maxwellian and non-Maxwellian collisional plasmas, we have made a survey of the X-ray spectral observations of stellar coronae to reveal the possibilities of such diagnostics in the detection of electron-energy distribution characteristics. Although several coronal high-resolution spectra have been previously reported with {\em Chandra} and {\em XMM-Newton} \citep{bri00, mew01, nes01, nes02b, nes03a, sch04, tes04a, leu07}, we make an attempt to improve the spectral fitting statistics with the {\sc xspec} package\footnote{https://heasarc.gsfc.nasa.gov/xanadu/xspec/}.

\begin{table}
  \caption{Observation list. \label{tab_obs}}
  \centering
  \begin{tabular}{lllrr} \hline\hline
  \sc{Source} & Observatory & Instrument & \# Obs.         \\ \hline
	Capella             & {\it Chandra}       & LETG--HRC  	&12    \\
		                 & {\it Chandra}       & HETG--ACIS  	&8    \\
		                 & {\it XMM--Newton}   & RGS    		&9    \\
	Procyon             & {\it Chandra}       & LETG--HRC  	&4    \\
		                 & {\it XMM--Newton}   & RGS    		&3    \\
	$\zeta$ Puppis      & {\it Chandra}       & LETG--HRC   	&1    \\
		                 & {\it Chandra}       & HETG--ACIS 	&1    \\
		                 & {\it XMM--Newton}   & RGS    		&9    \\
	$\beta$~Ceti        & {\it Chandra}       & LETG--HRC   	&1    \\
		                 & {\it Chandra}       & HETG--ACIS  	&1    \\
		                 & {\it XMM--Newton}   & RGS    		&1    \\
	$\lambda$~And       & {\it Chandra}       & LETG--HRC   	&1    \\
		                 & {\it Chandra}       & HETG--ACIS  	&1    \\
		                 & {\it XMM--Newton}   & RGS    		&1    \\
	AD~Leo              & {\it Chandra}       & LETG--HRC 	&2    \\
	$\xi$~Uma           & {\it Chandra}       & HETG--ACIS  	&1    \\
	Tz~CrB              & {\it Chandra}       & HETG--ACIS  	&1    \\
	AB~Dor              & {\it Chandra}       & HETG--ACIS  	&1    \\
	HR~1099             & {\it Chandra}       & LETG--HRC   	&1    \\
		                 & {\it Chandra}       & HETG--ACIS  	&1    \\
	V824-Ara            & {\it Chandra}       & HETG--ACIS  	&1    \\
	CC~Eri              & {\it Chandra}       & HETG--ACIS  	&2    \\
	$\mu$~Vel           & {\it Chandra}       & LETG--HRC   	&1    \\
		                 & {\it Chandra}       & HETG--ACIS  	&1    \\ \hline\hline
\end{tabular}
\end{table}

\begin{table}
  \caption{Fitted line emissivities ($10^{-18}$ photons cm$^{3}$\,s$^{-1}$\,bin$^{-1}$) for the He-like triplet of \ion{Ne}{ix} obtained with the {\sc vapec} collisional model of {\sc xspec}. \label{Ons11}}
  \vspace{0.3cm}
  \centering
  \renewcommand
  {\arraystretch}{1.2}
  \begin{tabular}{lcrrr}\hline\hline
  \sc{Source}    	& $k_BT$ (keV) & $r$ &  $i$     &  $f$    \\ \hline
	$\lambda$ And	& ${1.09}^{+0.11}_{-0.15}$ & $21.51$ 	& $2.62$ 	& $9.69$ 	\\
	$\mu$ Vel		& ${1.16}^{+0.35}_{-0.21}$ & $11.12$ 	& $1.28$ 	& $4.75$ 	\\
	$\xi$ Uma		& ${0.91}^{+0.09}_{-0.09}$ & $25.44$ 	& $3.10$ 	& $11.49$ 	\\
	$\beta$ Ceti	& ${1.12}^{+0.13}_{-0.09}$ & $15.05$ 	& $1.77$ 	& $6.56$ 	\\
	Capella			& ${1.09}^{+0.06}_{-0.07}$ & $15.05$ 	& $1.77$ 	& $6.56$ 	\\
	CC Eri			& ${0.45}^{+0.11}_{-0.06}$ & $289.61$ 	& $51.90$ 	& $184.83$ 	\\
	HR 1099			& ${0.99}^{+0.09}_{-0.11}$ & $25.44$ 	& $3.10$ 	& $11.49$ 	\\
	TZ CrB			& ${1.14}^{+0.04}_{-0.05}$ & $11.12$ 	& $1.28$ 	& $4.75$ 	\\
	V824 Ara			& ${1.24}^{+0.07}_{-0.08}$ & $7.19$ 	& $0.79$ 	& $2.90$ 	\\
  \hline\hline
  \end{tabular}
\end{table}

We have selected 13 bright sources from the Chandra Source Catalog (CSC)\footnote{http://cxc.harvard.edu/csc/} to analyse the He-like triplet spectral lines. Spectra were obtained from the Chandra Transmission Gratings Catalog and Archive (TGCat)\footnote{http://tgcat.mit.edu/} that include observations taken with both the High-Energy Transmission Grating (HETG) and the Low-Energy Transmission Grating (LETG) in combination with the Advanced CCD Imaging Spectrometer (ACIS) and the High Resolution Camera (HRC) detectors, respectively. In the case of Capella, Procyon, $\beta$~Ceti and $\lambda$~And, we considered observations obtained with {\it XMM-Newton} whose spectra were obtained with the Reflection Grating Spectrometer (RGS), and were reduced with the standard Scientific Analysis System (SAS) threads\footnote{http://xmm.esac.esa.int/sas}. The final sample consisted of a total of 65 observations that are listed in Table~\ref{tab_obs}. For each source, the available spectra were not combined but the fits were performed for each ionic species simultaneously by separately adjusting the spectral region centered on the corresponding He-like triplet: 40--41.8~\AA\ for \ion{C}{v}; 28.5--30~\AA\ for \ion{N}{vi}; 21.15--22.3~\AA\ for \ion{O}{vii}; 13.4--13.75~\AA\ for \ion{Ne}{ix}; 8.9--9.5~\AA\ for \ion{Mg}{xi} and 6.4--6.9~\AA\ for \ion{Si}{xiii}. For the {\sc xspec} spectral analysis, we fitted the emission-line region with a power-law continuum and Gaussian profiles, imposing a $C$-statistics variation of $\Delta C> 20$ to guarantee reliable line detection.

However, due to the complexity of the He-like triplet spectral region in \ion{Ne}{ix}, which may show blends with highly ionised Fe lines \citep[see, for instance,][]{mck80},  line emissivities rather than fluxes (see Table~\ref{Ons11}) were obtained by fitting it with the synthetic spectrum provided by the {\sc vapec} collisional model of {\sc xspec} based on the {\sc atomdb} (v3.0.2)\footnote{http://www.atomdb.org/} atomic database. It must be pointed out that for this ionic species {\sc atomdb} includes the effective collision strengths by \citet{che06} which, as discussed in Section~\ref{atomic_data}, have been found to be faulty. For the rest of the ions, namely \ion{Si}{xiii}, \ion{Mg}{xi}, \ion{O}{vii} and \ion{N}{vi}, the fitted line fluxes of our source sample are listed in Table~\ref{fluxes}, where the $i$-line MEG and HEG flux differences are due to instrumental effective areas. For $\zeta$ Puppis, we were not able to obtain fitted fluxes with Gaussian profiles which, as mentioned by \citet{leu07}, was apparently the result of resonant scattering effects in the He-like triplet spectral regions of \ion{N}{vi} and \ion{O}{vii}.

\begin{table*}
  \caption{Measured fluxes ($10^{-5}$ photons~cm$^{-2}$\,s$^{-1}$) for the $r$, $i$ and $f$ lines of the He-like triplet of \ion{Si}{xiii}, \ion{Mg}{xi}, \ion{O}{vii} and \ion{N}{vi} of our source sample. $^a$Observations with RGS. $^b$Observations with HEG. $^c$Observations with MEG. $^d$Observations with LETG. \label{fluxes}}
  \centering
  \renewcommand
  {\arraystretch}{1.4}
  \small
  \begin{tabular}{lrrrcrrrcrrr}\hline\hline
          & \multicolumn{3}{c}{\ion{Si}{xiii}} &    &  \multicolumn{3}{c}{\ion{Mg}{xi}} & & \multicolumn{3}{c}{\ion{O}{vii}} \\\cline{2-4}\cline{6-8}\cline{10-12}
{\sc Source} & $r$ & $i$ & $f$ &  &$r$ & $i$ & $f$ &  &$r$ & $i$ & $f$ \\ \hline
AB Dor			& $ {5.63^c}^{+0.64}_{-0.61} $ & $ {1.04^c}^{+0.38}_{-0.34} $ & $ {3.99^c}^{+0.56}_{-0.52} $ & & $ {4.35^c}^{+0.70}_{-0.64} $ & ... & $ {3.05^c}^{+0.61}_{-0.56} $ & & $ {28.36^c}^{+4.77}_{-4.68} $ & $ {10.74^c}^{+3.80}_{-3.82} $ & $ {17.81^c}^{+4.38}_{-4.14} $ \\
AD Leo			& $ {3.91^b}^{+0.89}_{-0.87} $ & $ {0.95^b}^{+0.49}_{-0.36} $ & $ {2.65^b}^{+0.63}_{-0.55} $ & & $ {1.94^c}^{+0.53}_{-0.46} $ & ... & $ {0.76^c}^{+0.39}_{-0.33} $ & & $ {47.00^c}^{+10.20}_{-9.15} $ & $ {13.45^c}^{+6.43}_{-5.18} $ & $ {24.57^c}^{+8.87}_{-7.63} $ \\
$\beta$ Ceti	& $ {8.71^b}^{+0.89}_{-0.84} $ & $ {2.27^b}^{+0.50}_{-0.45} $ & $ {7.40^b}^{+0.76}_{-0.72} $ & & $ {11.14^b}^{+1.16}_{-1.09} $ & $ {2.88^b}^{+0.85}_{-0.76} $ & $ {6.77^b}^{+0.98}_{-0.91} $ & & $ {14.51^c}^{+3.96}_{-3.64} $ & ... & $ {11.07^c}^{+3.65}_{-4.19} $ \\
					& $ {9.40^c}^{+0.60}_{-0.57} $ & $ {2.34^c}^{+0.34}_{-0.32} $ & $ {6.50^c}^{+0.49}_{-0.47} $ & & $ {12.64^c}^{+0.82}_{-0.79} $ & $ {2.47^c}^{+0.44}_{-0.41} $ & $ {6.75^c}^{+0.67}_{-0.63} $ & & $ {11.76^d}^{+2.53}_{-2.02} $ & ... & $ {8.78^d}^{+2.27}_{-2.02} $ \\
CC Eri	& $ {4.04^b}^{+0.63}_{-0.57} $ & $ {0.84^b}^{+0.33}_{-0.27} $ & $ {2.33^b}^{+0.44}_{-0.39} $ & & $ {1.77^b}^{+0.50}_{-0.43} $ & ... & $ {0.72^b}^{+0.45}_{-0.35} $ & & $ {52.97^c}^{+8.52}_{-7.70} $ & $ {13.80^c}^{+4.77}_{-4.05} $ & $ {25.89^c}^{+6.91}_{-5.86} $ \\
	& $ {4.27^c}^{+0.41}_{-0.39} $ & $ {1.06^c}^{+0.29}_{-0.25} $ & $ {2.78^c}^{+0.33}_{-0.31} $ & & $ {3.01^c}^{+0.42}_{-0.40} $ & $ {0.54^c}^{+0.30}_{-0.25} $ & $ {1.45^c}^{+0.32}_{-0.29} $ & & ... & ... & ... \\
HR 1099	& $ {10.03^b}^{+1.10}_{-1.03} $ & $ {1.79^b}^{+0.63}_{-0.58} $ & $ {8.15^b}^{+0.94}_{-0.89} $ & & $ {9.37^b}^{+1.17}_{-1.05} $ & $ {1.78^b}^{+1.68}_{-0.76} $ & $ {4.33^b}^{+0.89}_{-0.83} $ & & $ {39.87^c}^{+5.73}_{-5.33} $ & $ {6.32^c}^{+3.35}_{-2.79} $ & $ {22.36^c}^{+5.20}_{-4.68} $ \\
	& $ {12.11^c}^{+0.78}_{-0.76} $ & $ {2.00^c}^{+0.53}_{-0.50} $ & $ {7.71^c}^{+0.67}_{-0.65} $ & & $ {10.54^c}^{+0.81}_{-0.78} $ & $ {2.13^c}^{+0.83}_{-0.76} $ & $ {3.88^c}^{+0.65}_{-0.63} $ & & $ {34.43^d}^{+3.50}_{-3.37} $ & $ {9.11^d}^{+2.62}_{-2.48} $ & $ {21.32^d}^{+3.18}_{-3.05} $ \\
$\lambda$ And	& $ {7.07^b}^{+0.88}_{-0.81} $ & $ {1.24^b}^{+0.47}_{-0.43} $ & $ {4.55^b}^{+0.62}_{-0.58} $ & & $ {11.18^b}^{+1.15}_{-1.09} $ & $ {0.96^b}^{+0.48}_{-0.41} $ & $ {4.45^b}^{+0.82}_{-0.73} $ & & $ {17.77^c}^{+4.29}_{-3.73} $ & ... & $ {9.93^c}^{+3.57}_{-3.12} $ \\
	& $ {7.70^c}^{+0.59}_{-0.56} $ & $ {0.98^c}^{+0.29}_{-0.27} $ & $ {4.47^c}^{+0.46}_{-0.45} $ & & $ {11.23^c}^{+0.74}_{-0.72} $ & $ {2.33^c}^{+0.53}_{-0.48} $ & $ {5.17^c}^{+0.56}_{-0.53} $ & & $ {25.17^d}^{+4.89}_{-4.61} $ & ... & $ {17.86^d}^{+4.64}_{-4.29} $ \\
$\mu$ Vel	& $ {3.99^b}^{+0.75}_{-0.67} $ & $ {1.33^b}^{+0.47}_{-0.40} $ & $ {2.13^b}^{+0.51}_{-0.44} $ & & $ {5.85^b}^{+1.06}_{-1.12} $ & $ {1.45^b}^{+0.62}_{-0.56} $ & $ {3.52^b}^{+0.83}_{-0.73} $ & & ... & ... & ... \\
	& $ {3.63^c}^{+0.47}_{-0.43} $ & $ {1.15^c}^{+0.29}_{-0.26} $ & $ {2.52^c}^{+0.38}_{-0.35} $ & & $ {5.30^c}^{+0.61}_{-0.57} $ & $ {0.95^c}^{+0.35}_{-0.31} $ & $ {2.66^c}^{+0.50}_{-0.44} $ & & ... & ... & ... \\
TZ CrB	& $ {13.32^b}^{+1.18}_{-1.12} $ & $ {1.95^b}^{+0.58}_{-0.52} $ & $ {8.33^b}^{+0.91}_{-0.86} $ & & $ {16.47^b}^{+1.49}_{-1.42} $ & $ {3.17^b}^{+0.79}_{-0.71} $ & $ {7.42^b}^{+1.07}_{-0.99} $ & & $ {34.19^c}^{+6.05}_{-5.50} $ & $ {10.45^c}^{+3.96}_{-3.43} $ & $ {26.06^c}^{+5.72}_{-5.01} $ \\
	& $ {14.27^c}^{+0.81}_{-0.78} $ & $ {2.37^c}^{+0.43}_{-0.41} $ & $ {8.59^c}^{+0.65}_{-0.63} $ & & $ {18.17^c}^{+1.34}_{-1.29} $ & $ {2.93^c}^{+0.68}_{-0.63} $ & $ {7.62^c}^{+0.82}_{-0.79} $ & & ... & ... & ... \\
Procyon	& ... & ... & ... & & ... & ... & ... & & $ {38.38^d}^{+4.29}_{-4.07} $ & $ {8.94^d}^{+2.44}_{-2.17} $ & $ {35.12^d}^{+4.14}_{-3.89} $ \\
V824 Ara	& $ {5.10^b}^{+0.71}_{-0.66} $ & $ {0.83^b}^{+0.38}_{-0.32} $ & $ {3.67^b}^{+0.59}_{-0.55} $ & & $ {4.38^b}^{+0.80}_{-0.73} $ & ... & $ {2.97^b}^{+0.69}_{-0.60} $ & & $ {19.98^c}^{+5.01}_{-4.30} $ & $ {6.89^c}^{+3.39}_{-2.71} $ & $ {10.45^c}^{+4.35}_{-3.56} $ \\
	& $ {5.54^c}^{+0.49}_{-0.47} $ & $ {1.21^c}^{+0.32}_{-0.30} $ & $ {3.20^c}^{+0.38}_{-0.36} $ & & $ {5.88^c}^{+0.65}_{-0.62} $ & $ {1.08^c}^{+0.47}_{-0.40} $ & $ {2.76^c}^{+0.50}_{-0.47} $ & & ... & ... & ... \\
$\xi$ Uma	& $ {4.49^b}^{+0.73}_{-0.70} $ & $ {1.02^b}^{+0.46}_{-0.36} $ & $ {3.02^b}^{+0.57}_{-0.51} $ & & $ {7.45^b}^{+1.05}_{-0.97} $ & $ {1.36^b}^{+0.55}_{-0.45} $ & $ {5.04^b}^{+0.90}_{-0.82} $ & & $ {55.84^c}^{+7.83}_{-7.17} $ & $ {14.64^c}^{+4.38}_{-3.66} $ & $ {30.11^c}^{+6.74}_{-6.20} $ \\
	& $ {5.15^c}^{+0.50}_{-0.47} $ & $ {0.89^c}^{+0.26}_{-0.23} $ & $ {3.02^c}^{+0.39}_{-0.36} $ & & $ {7.96^c}^{+0.68}_{-0.65} $ & $ {1.67^c}^{+0.38}_{-0.35} $ & $ {4.02^c}^{+0.52}_{-0.49} $ & & ... & ... & ... \\
Capella	& $ {27.59^b}^{+2.77}_{-2.60} $ & $ {5.93^b}^{+1.48}_{-1.28} $ & $ {20.07^b}^{+2.20}_{-2.06} $ & & $ {40.40^b}^{+3.47}_{-3.29} $ & $ {7.35^b}^{+1.87}_{-1.60} $ & $ {20.68^b}^{+2.66}_{-2.47} $ & & $ {118.5^c}^{+18.76}_{-17.21} $ & $ {25.86^c}^{+11.26}_{-9.09} $ & $ {73.96^c}^{+17.55}_{-15.01} $ \\
			& $ {28.37^c}^{+1.74}_{-1.68} $ & $ {6.58^c}^{+0.96}_{-0.89} $ & $ {19.85^c}^{+1.44}_{-1.38} $ & & $ {43.65^c}^{+2.38}_{-2.31} $ & $ {9.07^c}^{+1.41}_{-1.30} $ & $ {22.15^c}^{+1.93}_{-1.84} $ & & $ {90.82^d}^{+11.60}_{-10.76} $ & $ {15.88^d}^{+6.61}_{-5.51} $ & $ {65.41^d}^{+11.08}_{-10.04} $ \\
			&				...			&		...			&			...		& &			...				&		...			&			...		& & $ {115.64^a}^{+4.63}_{-4.69} $ & $ {16.17^a}^{+2.77}_{-2.64} $ & $ {82.50^a}^{+4.13}_{-4.02} $ \\ \hline
          &    & N {\sc vi} &    &    &  &   & & & &  & \\\cline{2-4}
{\sc Source} & $r$ & $i$ & $f$ &  & &  &  &  & &  &  \\\cline{1-4}
Procyon      & ${11.19^a}^{+1.62}_{-2.30}$ & ${4.39^a}^{+1.08}_{-1.59}$ & ${6.10^a}^{+1.46}_{-1.46}$ & & & & & & \\
Capella      & ${21.17^a}^{+3.77}_{-3.53}$ & ${8.11^a}^{+3.05}_{-2.86}$ & ${24.54^a}^{+4.34}_{-4.12}$ & & & & & & \\ \hline\hline
\end{tabular}
\label{Table2}
\end{table*}

\begin{table*}
  \caption{\ion{O}{vii} $R(N_e)$ and $G(T_e)$ line ratios extracted from the {\em Chandra} MEG spectra of seven sources and the corresponding derived electron temperatures. The MEG and LETGS results by \citet{tes04a} and \citet{nes02b}, respectively, are also included. \label{Diag-O}}
  \centering
  \small
  \renewcommand
{\arraystretch}{1.2}
\begin{tabular}{lccccccccccc}\hline\hline
					&	\multicolumn{3}{c}{Present Work} && \multicolumn{2}{c}{\citet{tes04a}} && \multicolumn{3}{c}{\citet{nes02b}} \\ \cline{2-4} \cline{6-7} \cline{9-11}
\sc{Source} & $G(T_e)$					 & $R(N_e)$ 	 			& $T_e$(MK) &&	$G(T_e)$		& $R(N_e)$	&&	$G(T_e)$		& $R(N_e)$	& $T_e$(MK)	  \\ \hline
AB Dor		& $ 1.01^{+0.27}_{-0.26} $ & $ 1.66^{+0.72}_{-0.71} $ & $1.26^{+0.22}_{-0.19}$ && $ 1.00 \pm 0.28$ & $ 1.76 \pm 0.86 $ 	& & ... 	& ... & ... \\
AD Leo		& $ 0.81^{+0.29}_{-0.25} $ & $ 1.83^{+1.09}_{-0.90} $ & $1.86^{+0.28}_{-0.24}$ && ... 					& ... 					& & $ 0.86 \pm 0.11$ & $ 3.22 \pm 0.62	$ 	& $2.2 \pm 0.6$\\
HR 1099		& $ 0.88^{+0.15}_{-0.14} $ & $ 2.34^{+0.76}_{-0.72} $ & $1.91^{+0.33}_{-0.28}$ && $ 0.81 \pm 0.15$ & $ 2.89 \pm 1.06 $ 	& & $ 0.80 \pm 0.10$ & $ 2.22 \pm 0.41	$ 	& $2.52 \pm 0.64$\\
Capella		& $ 0.90^{+0.18}_{-0.16} $ & $ 4.12^{+1.85}_{-1.56} $ & $1.82^{+0.32}_{-0.26}$ && ... 					& ... 					& & $ 0.90 \pm 0.03$ & $ 3.92 \pm 0.25	$ 	& $2.0 \pm 0.13$\\
TZ CrB		& $ 1.07^{+0.28}_{-0.25} $ & $ 2.49^{+1.09}_{-0.95} $ & $1.15^{+0.17}_{-0.19}$ && $ 1.18 \pm 0.24$ & $ 2.32 \pm 0.75 $ 	& & ...					& ...	& ...  \\
Procyon		& $ 1.15^{+0.18}_{-0.17} $ & $ 3.93^{+1.17}_{-1.05} $ & $1.00^{+0.17}_{-0.17}$ && ... 					& ... 					& & $ 1.21 \pm 0.08$ & $ 3.22 \pm 0.3$ & $0.96 \pm 0.25$\\
$\xi$ Uma	& $ 0.80^{+0.18}_{-0.16} $ & $ 2.06^{+0.77}_{-0.67} $ & $1.91^{+0.28}_{-0.28}$ && $ 0.79 \pm 0.16$ & $ 2.38 \pm 0.95 $ 	& & ... 					& ...	& ...	\\
\hline\hline
\end{tabular}
\end{table*}

\begin{table*}
  \caption{Comparison of the present \ion{Si}{xiii} and \ion{Mg}{xi} line ratios with \citet{tes04a}. For each source, the first and second lines respectively list the {\em Chandra} HEG and MEG values. \label{Diag-Si}}
  \centering
  \small
  \renewcommand
{\arraystretch}{1.3}
\begin{tabular}{lccccccccccc}\hline\hline
					&	\multicolumn{5}{c}{\ion{Si}{xiii}}& &\multicolumn{5}{c}{\ion{Mg}{xi}} \\ \cline{2-6} \cline{8-12}
					&	\multicolumn{3}{c}{Present Work} & \multicolumn{2}{c}{\citet{tes04a}} &	&\multicolumn{3}{c}{Present Work} & \multicolumn{2}{c}{\citet{tes04a}} \\
\sc{Source} 	& $G(T_e)$					 	 & $R(N_e)$ 	 						&	$T_e$(MK)	  	 &	$G(T_e)$			  & $R(N_e)$&	& $G(T_e)$					 	 & $R(N_e)$ 	 						&	$T_e$(MK)	  	 						&	$G(T_e)$			  & $R(N_e)$	  	 \\ \hline
AB Dor & ... & ... & ... & $ 1.04 \pm 0.18 $ & $ 3.46 \pm 1.39                $  & & ...   & ...   & ...     &  ...     & ... \\
  & $ 0.89 ^{+ 0.16}_{-0.15 } $ & $ 3.85 ^{+ 1.51}_{-1.37 } $ & $ 4.90 ^{+ 0.99}_{-0.92 } $ & $ 0.87 \pm 0.11 $ & $ 3.37 \pm 0.88 $  & &...   & ...   & ... &  ...     & ...     \\
$\beta$ Ceti & $ 1.11 ^{+ 0.15}_{-0.14 } $ & $ 3.25 ^{+ 0.79}_{-0.71 } $ & $ 2.75 ^{+ 0.71}_{-0.66 } $ & $ 1.07 \pm 0.11 $ & $ 3.14 \pm 0.51 $  & & $ 0.87^{+0.15}_{-0.14} $ & $ 2.35^{+0.77}_{-0.70} $ &  $ 3.98 ^{+ 0.70 }_{ -0.75 } $ &  $ 0.92 \pm 0.10 $ & $ 3.00 \pm 0.64 $ \\
  & $ 0.94 ^{+ 0.09}_{-0.08 } $ & $ 2.78 ^{+ 0.46}_{-0.43 } $ & $ 4.37 ^{+ 0.88}_{-0.90 } $ & $ 0.87 \pm 0.05 $ & $ 2.42 \pm 0.28 $  & & $ 0.73^{+0.08}_{-0.08} $ & $ 2.74^{+0.56}_{-0.53} $ &  $ 5.50 ^{+ 0.81 }_{ -0.92 } $ &  $ 0.86 \pm 0.02 $ & $ 2.25 \pm 0.10 $ \\
HR 1099 & $ 0.99 ^{+ 0.16}_{-0.15 } $ & $ 4.55 ^{+ 1.69}_{-1.56 } $ & $ 3.80 ^{+ 0.88}_{-0.78 } $ & $ 1.06 \pm 0.19 $ & $ 2.40 \pm 0.78 $  & & $ 0.65^{+0.22}_{-0.14} $ & $ 2.43^{+2.35}_{-1.14} $ &  $ 6.76 ^{+ 1.00 }_{ -1.14 } $ &  $ 0.75 \pm 0.10 $ & $ 2.12 \pm 0.56 $ \\
  & $ 0.80 ^{+ 0.09}_{-0.08 } $ & $ 3.86 ^{+ 1.07}_{-1.03 } $ & $ 6.17 ^{+ 1.08}_{-1.04 } $ & $ 1.09 \pm 0.15 $ & $ 1.79 \pm 0.39 $  & & $ 0.57^{+0.11}_{-0.10} $ & $ 1.82^{+0.77}_{-0.71} $ &  $ 8.32 ^{+ 1.23 }_{ -1.24 } $ &  $ 0.77 \pm 0.04 $ & $ 2.26 \pm 0.21 $ \\
$\lambda$ And & $ 0.82 ^{+ 0.15}_{-0.14 } $ & $ 3.68 ^{+ 1.49}_{-1.36 } $ & $ 5.89 ^{+ 1.03}_{-0.99 } $ & $ 0.90 \pm 0.09 $ & $ 2.93 \pm 0.69 $  & & $ 0.48^{+0.10}_{-0.09} $ & $ 4.65^{+2.47}_{-2.12} $ &  $ 11.22 ^{+ 1.96 }_{ -1.67 } $ &  $ 0.65 \pm 0.08 $ & $ 3.00 \pm 0.87 $ \\
  & $ 0.71 ^{+ 0.09}_{-0.09 } $ & $ 4.57 ^{+ 1.44}_{-1.33 } $ & $ 7.94 ^{+ 1.39}_{-1.34 } $ & $ 0.68 \pm 0.06 $ & $ 4.16 \pm 0.97 $  & & $ 0.67^{+0.08}_{-0.08} $ & $ 2.22^{+0.56}_{-0.51} $ &  $ 6.31 ^{+ 1.10 }_{ -0.94 } $ &  $ 0.68 \pm 0.01 $ & $ 2.40 \pm 0.11 $ \\
$\mu$ Vel & $ 0.87 ^{+ 0.24}_{-0.21 } $ & $ 1.59 ^{+ 0.68}_{-0.58 } $ & $ 5.25 ^{+ 0.92}_{-0.98 } $ & $ 0.91 \pm 0.21 $ & $ 2.00 \pm 0.81 $  & & $ 0.85^{+0.23}_{-0.22} $ & $ 2.43^{+1.18}_{-1.06} $ &  $ 4.17 ^{+ 0.73 }_{ -0.78 } $ &  $ 0.72 \pm 0.16 $ & $ 3.10 \pm 1.66 $ \\
  & $ 1.01 ^{+ 0.19}_{-0.17 } $ & $ 2.19 ^{+ 0.65}_{-0.58 } $ & $ 3.63 ^{+ 0.84}_{-0.81 } $ & $ 0.96 \pm 0.11 $ & $ 1.95 \pm 0.39 $  & & $ 0.68^{+0.14}_{-0.12} $ & $ 2.81^{+1.17}_{-1.02} $ &  $ 6.17 ^{+ 1.08 }_{ -0.92 } $ &  $ 0.70 \pm 0.04 $ & $ 3.14 \pm 0.51 $ \\
TZ CrB & $ 0.77 ^{+ 0.11}_{-0.10 } $ & $ 4.27 ^{+ 1.35}_{-1.21 } $ & $ 6.76 ^{+ 1.18}_{-1.14 } $ & $ 0.75 \pm 0.07 $ & $ 3.60 \pm 0.76 $  & & $ 0.64^{+0.10}_{-0.09} $ & $ 2.34^{+0.67}_{-0.61} $ &  $ 6.92 ^{+ 1.02 }_{ -1.03 } $ &  $ 0.76 \pm 0.07 $ & $ 1.98 \pm 0.32 $ \\
  & $ 0.77 ^{+ 0.07}_{-0.07 } $ & $ 3.62 ^{+ 0.71}_{-0.67 } $ & $ 6.76 ^{+ 1.18}_{-1.14 } $ & $ 0.76 \pm 0.04 $ & $ 2.92 \pm 0.35 $  & & $ 0.58^{+0.07}_{-0.07} $ & $ 2.60^{+0.66}_{-0.62} $ &  $ 8.13 ^{+ 1.20 }_{ -1.21 } $ &  $ 0.73 \pm 0.03 $ & $ 2.36 \pm 0.21 $ \\
$\xi$ Uma & $ 0.90 ^{+ 0.22}_{-0.20 } $ & $ 2.97 ^{+ 1.44}_{-1.17 } $ & $ 4.79 ^{+ 0.97}_{-0.90 } $ & $ 0.98 \pm 0.18 $ & $ 2.08 \pm 0.71 $  & & $ 0.86^{+0.19}_{-0.17} $ & $ 3.69^{+1.62}_{-1.37} $ &  $ 4.07 ^{+ 0.71 }_{ -0.76 } $ &  $ 0.97 \pm 0.14 $ & $ 2.89 \pm 0.85 $ \\
  & $ 0.76 ^{+ 0.12}_{-0.11 } $ & $ 3.40 ^{+ 1.08}_{-0.97 } $ & $ 6.92 ^{+ 1.21}_{-1.16 } $ & $ 0.72 \pm 0.08 $ & $ 2.94 \pm 0.68 $  & & $ 0.71^{+0.10}_{-0.10} $ & $ 2.41^{+0.63}_{-0.58} $ &  $ 5.75 ^{+ 0.85 }_{ -0.86 } $ &  $ 0.90 \pm 0.04 $ & $ 2.27 \pm 0.13 $ \\
\hline\hline
\end{tabular}
\end{table*}

The present \ion{O}{vii} $G(T_e)$ and $R(N_e)$ line ratios obtained from the {\em Chandra} MEG spectra of seven sources and our derived electron temperatures are compared in Table~\ref{Diag-O} with the results by \citet{nes02b} and \citet{tes04a}. The $G(T_e)$ ratios agree to $\sim 10\%$ (within the error bars), but larger differences ($\sim 20\%$) are found for $R(N_e)$; for AD Leo, the difference with \citet{nes02b} is almost a factor of 2. Regarding the latter diagnostic, error bars by \citet{nes02b} are somewhat smaller than \citet{tes04a} and the present. The electron temperature accord (around 10\%), on the other hand, is found to be satisfactory.

\begin{figure*}
  \centering
  \begin{tabular}{cc}
  \includegraphics[scale=0.40]{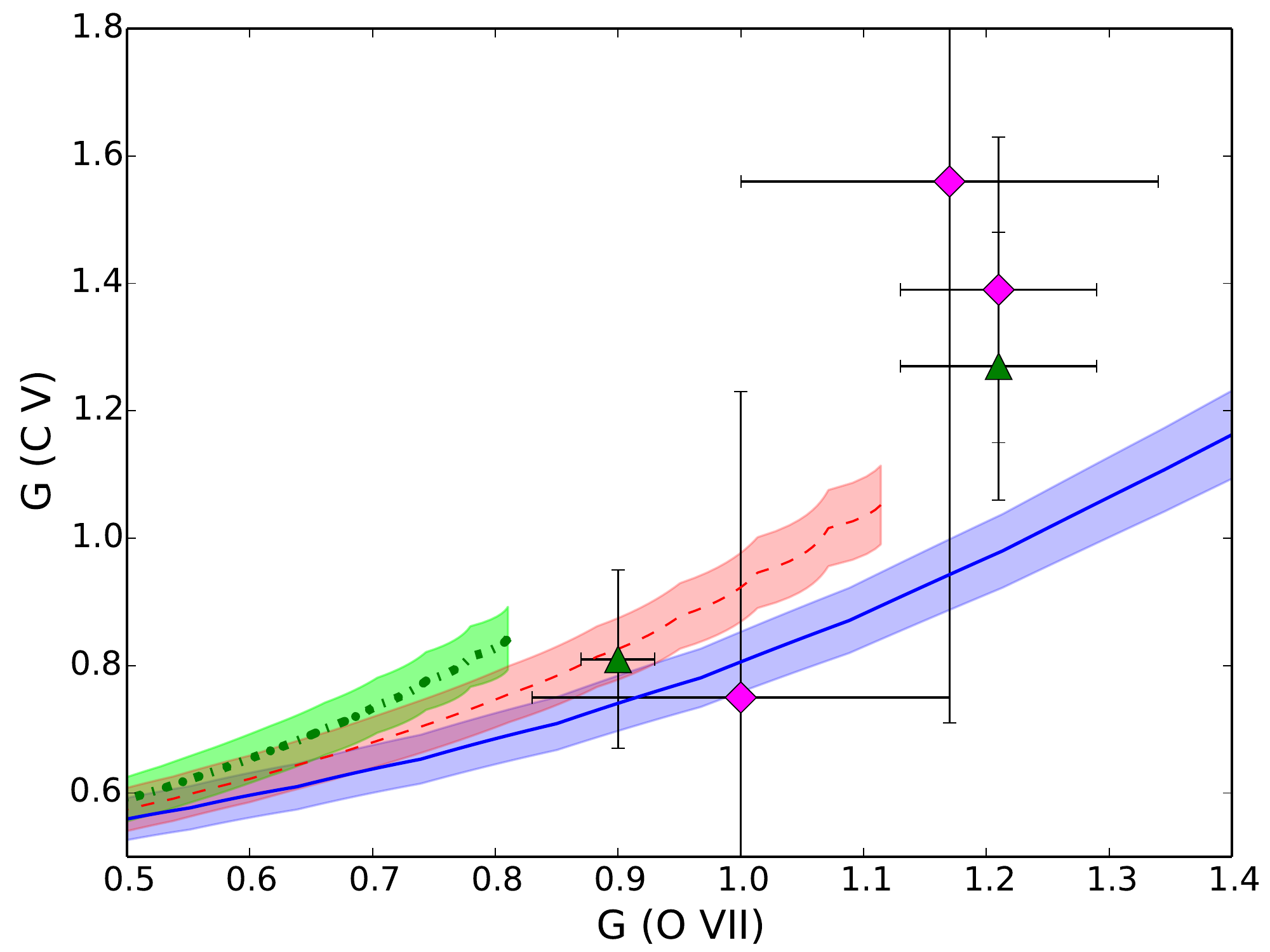} &
  \includegraphics[scale=0.40]{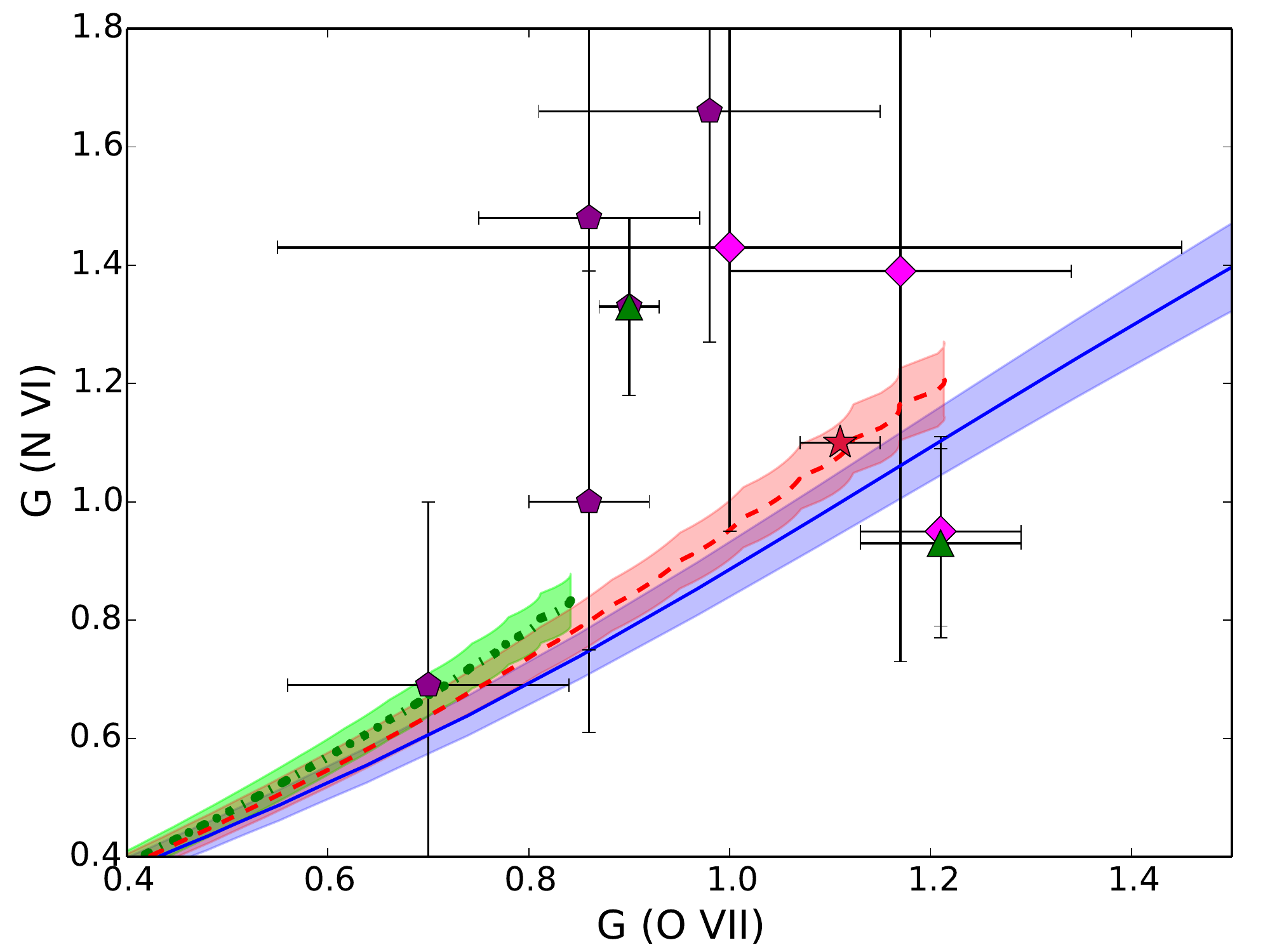} \\
  \includegraphics[scale=0.40]{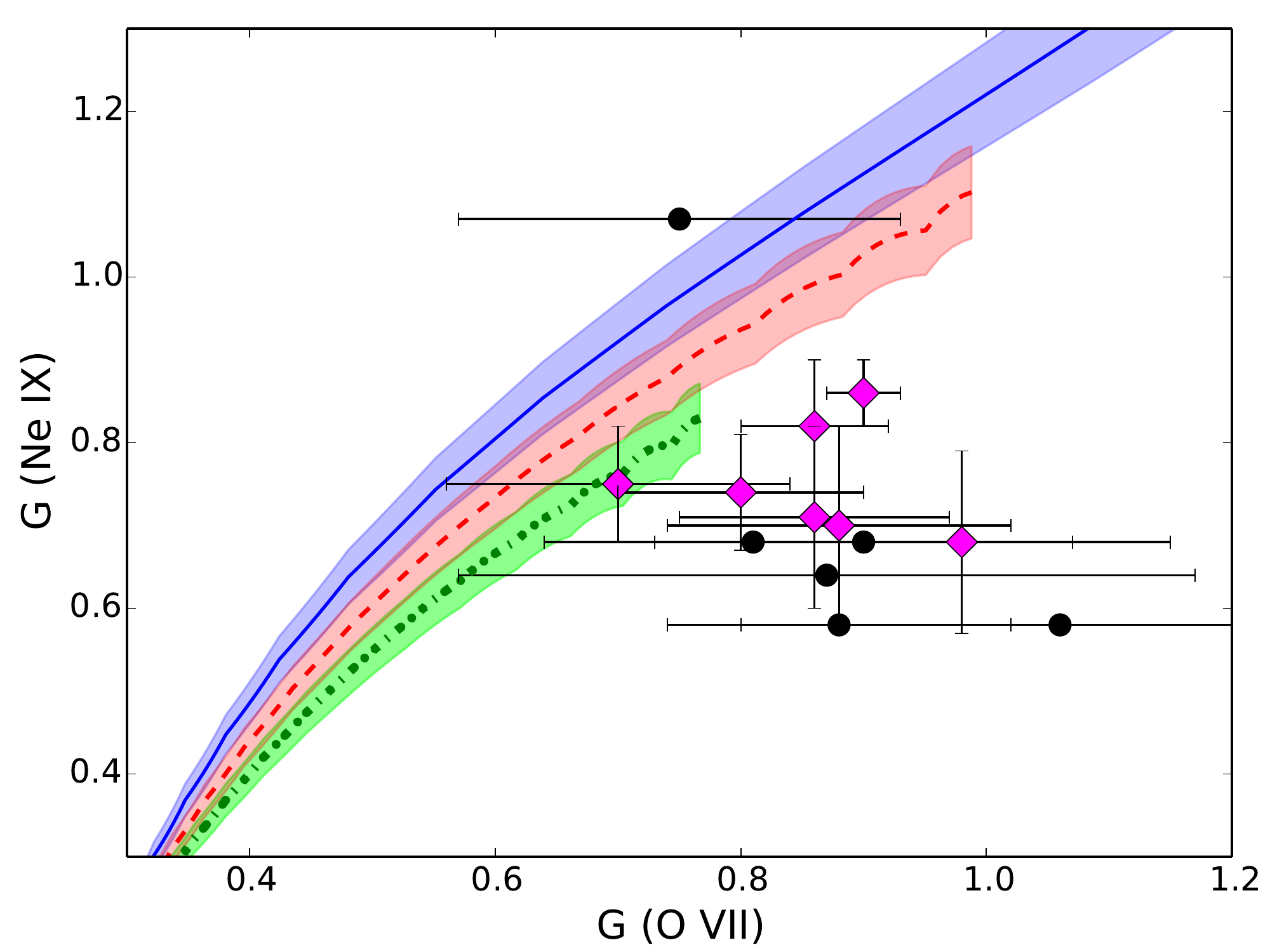} &
  \includegraphics[scale=0.40]{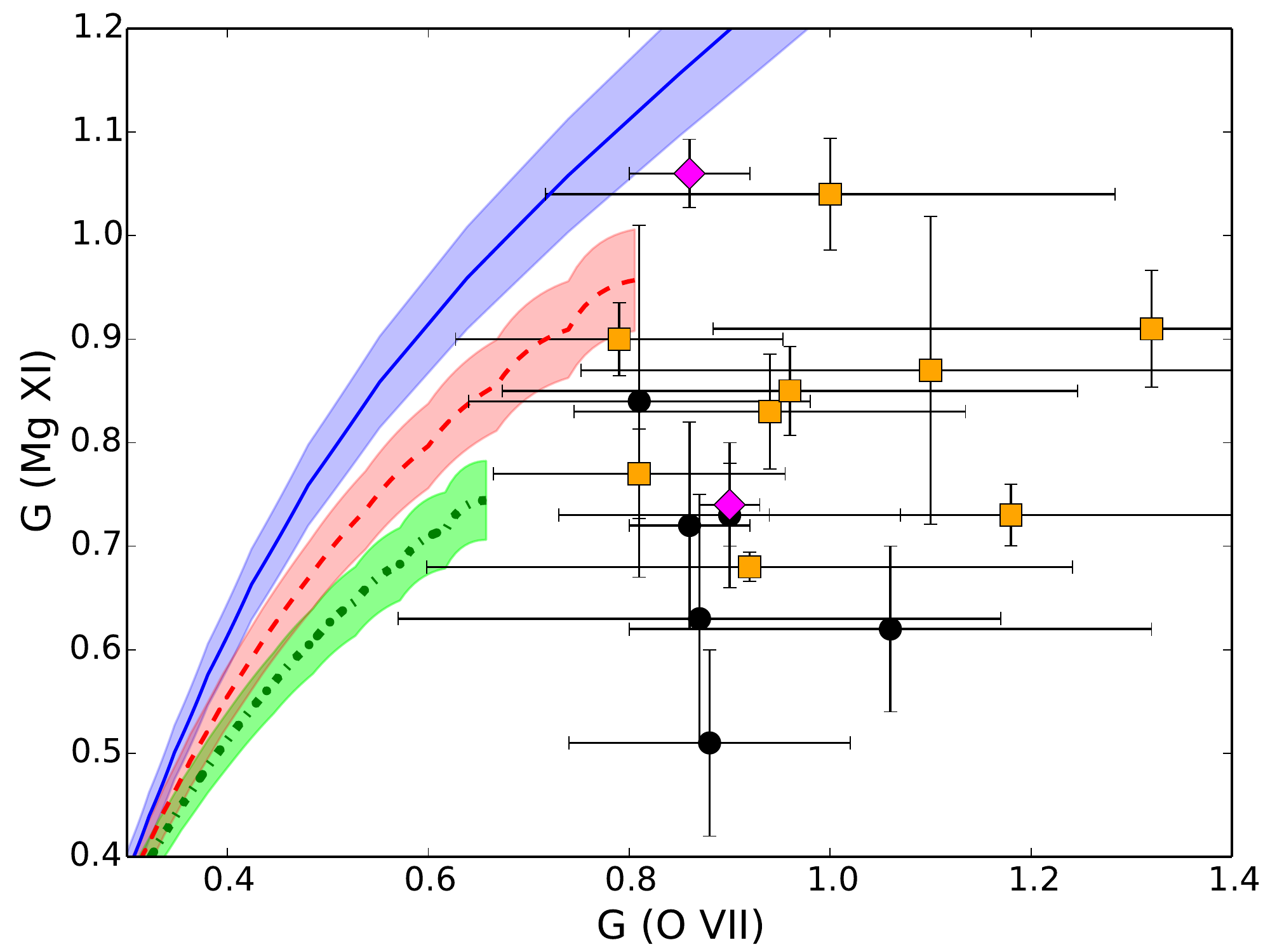} \\
  \includegraphics[scale=0.40]{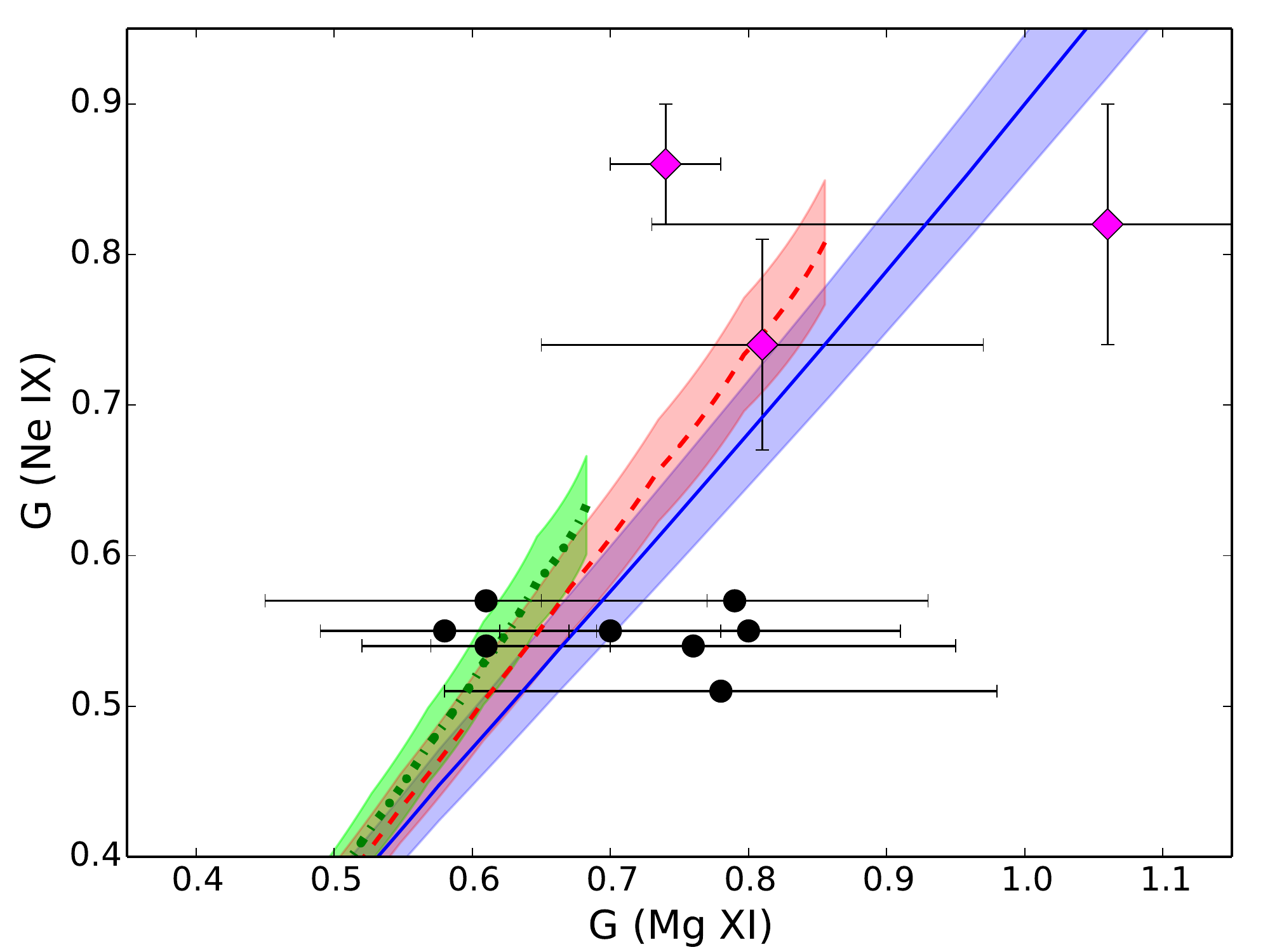} &
  \includegraphics[scale=0.40]{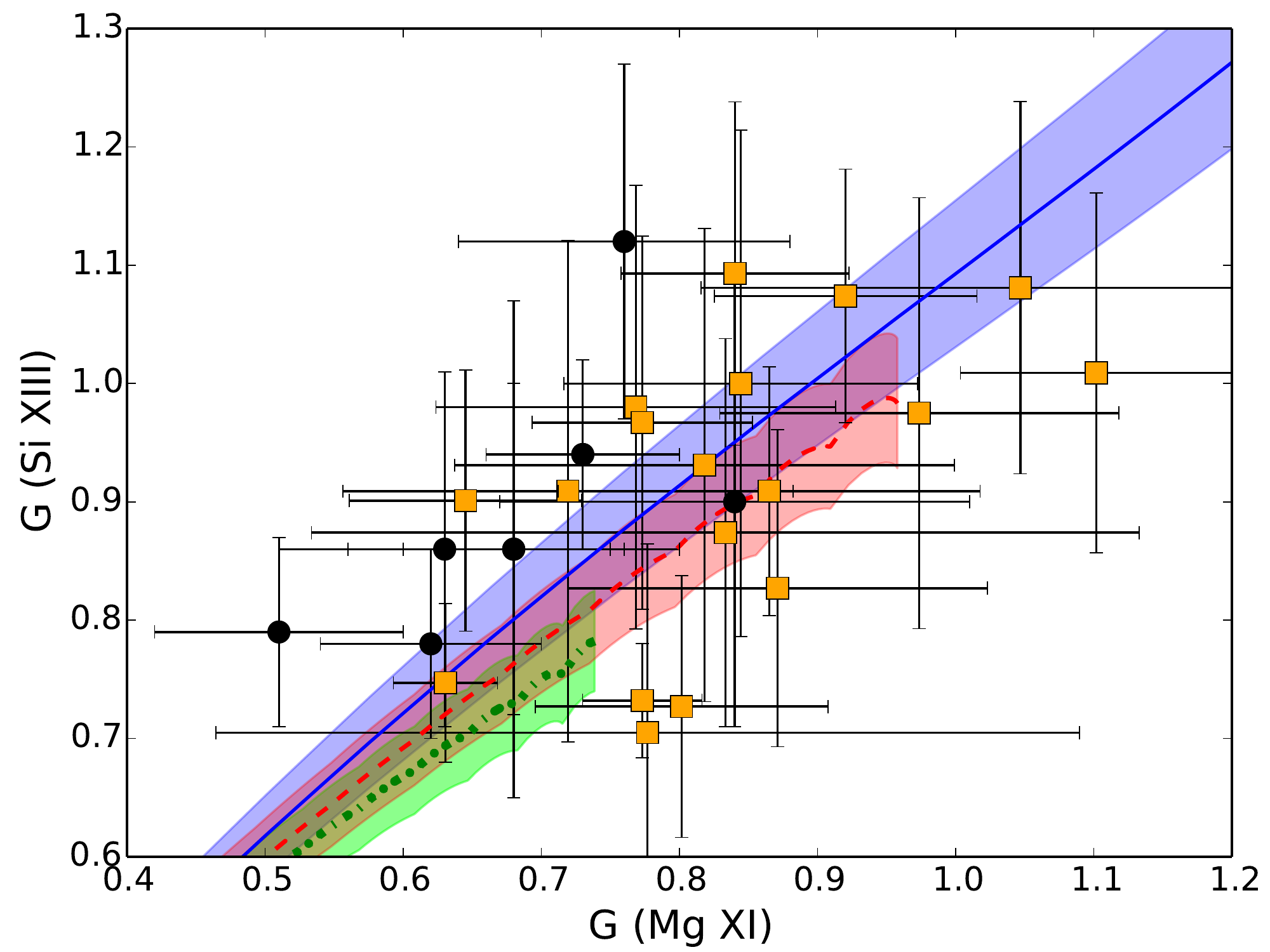} \\
  \end{tabular}
  \caption{Comparison of observed (stellar coronae) $G(T_e)$ line ratios and the theoretical loci for \ion{C}{v}, \ion{N}{vi}, \ion{O}{vii}, \ion{Ne}{ix}, \ion{Mg}{xi} and \ion{Si}{xiii}. Black circles: present data. Yellow squares: \citet{tes04a}. Purple diamonds: \citet{nes02b}. Green triangles: \citet{nes01}. Red star:  \citet{leu07}. The continuous blue curve assumes an MB distribution while the red dashed and green dash-dotted curves consider $\kappa$ distributions with $\kappa=10$ and $\kappa=5$, respectively.  \label{benchmark}}
\end{figure*}

In a similar fashion, a comparison of the present \ion{Si}{xiii} and \ion{Mg}{xi} line ratios with those by \citet{tes04a} in seven sources is given in Table~\ref{Diag-Si}. For \ion{Si}{xiii}, it may be appreciated that, in general, the MEG $G(T_e)$ line ratio is 15\% smaller that the HEG value while $R(N_e)$ is 20\% larger; the latter difference is mainly due to underestimates of the MEG $i$-line flux. Good agreement is found between the present \ion{Si}{xiii} line ratios and \citet{tes04a}. For \ion{Mg}{xi}, the same aforementioned MEG--HEG pattern is observed, but larger line-ratio differences (25\%) are found with respect to  \citet{tes04a} which, as previously mentioned by these authors, are probably due to blending with lines from the \ion{Ne}{x} Lyman series and highly ionised Fe.

It must be mentioned that blending of the He-like triplet components with highly ionised Fe lines, such as that reported in the \ion{Ne}{xi} and \ion{Mg}{xi} spectral regions, would be sensitive to flares \citep{mck80, tes04a}. In the present spectral reductions the light curve of each spectrum was carefully analysed during the whole duration of the observation in order to detect variations that could be assigned to flaring events. However, for most of the sources that have been considered---namely $\beta$~Ceti, Procyon, $\xi$~Uma, $\mu$~Vel, AD~Leo, $\lambda$~And and Capella---no significant variations were found or they were so short lasted as not to affect the total spectrum. However, two LETG Capella spectra (1420 and 62435) out of 12 showed a flare with an approximate duration of 8\,Ks at the beginning of the first and end of the second observations, and were thus not taken into account. CC~Eri experimented a flare at the end of the observation that increased the emitted flux by a factor of 6; a similar situation was observed in TZ~CrB where the emission increased by a factor of 4.5 during the last 3\,Ks of the total observing time of 83.7\,Ks; and the emission flux of AB~Dor was practically constant except for a flare lasting 10\,Ks during the 54\,Ks observing time.

Regarding the two \ion{N}{vi} observations in Capella and Procyon (see Table~\ref{fluxes}), we find very good agreement ($\sim 5\%$) with the values of $G(T_e)$, $R(N_e)$ and $T_e$ listed by \citet{nes01, nes02b}. However, we do not list $r$, $i$ and $f$ fluxes for \ion{C}{v} since the Gaussian fits in the available spectra did not fulfill our prefixed statistical expectations.

In Fig.~\ref{benchmark}, temperature diagnostic maps are shown for the ions considered in the present study that include the observed line ratios and the theoretical loci for both MB and $\kappa$ distributions. There are two facts that clearly stand out: (i) the differences in the theoretical loci caused by the electron-energy distributions (i.e. MB {\em vs}. $\kappa$) are generally small and are mostly overlapped by the uncertainty margins due to the atomic data errors; and (ii) the observational error bars are far too large to be able to discern the electron-energy distribution type or even to firmly deduce plasma ionic coexistence. Nonetheless, these diagnostic maps appear to indicate stellar coronae composed of at least two plasma regions: a hot component ($T_e\gtrsim 6$~MK) where \ion{Ne}{ix}, \ion{Mg}{xi} and \ion{Si}{xiii} coexist; and a cooler region ($T_e\lesssim 2$~MK) containing \ion{C}{v}, \ion{N}{vi} and \ion{O}{vii}.

\begin{table*}
  \caption{Comparison of the present $R(N_e)$ and $G(T_e)$ line ratios in \ion{Mg}{xi} with those measured in a tokamak experiment and previously calculated \citep{kee91}.$^a$ \label{tokamak_mg}}
  \centering
  \small
  \renewcommand
  {\arraystretch}{1.3}
  \begin{tabular}{cccccccccc}
  \hline\hline
  & & \multicolumn{2}{c}{Present work} && \multicolumn{2}{c}{Expt$^a$} && \multicolumn{2}{c}{Theory$^a$}\\
  \cline{3-4} \cline{6-7}\cline{9-10}
  $N_e(10^{13}$\,cm$^{-3}$) & $T_e(10^2$\,eV) & $R(N_e)$ & $G(T_e)$ && $R(N_e)$ & $G(T_e)$ && $R(N_e)$ & $G(T_e)$ \\
  \hline
  1.1 & 7.0 & $1.25\pm 0.05$ & $0.62\pm 0.03$ && 1.45 & 0.73 && 1.40 & 0.71 \\
  1.7 & 7.0 & $0.96\pm 0.04$ & $0.62\pm 0.03$ && 1.04 & 0.81 && 1.07 & 0.71 \\
  2.2 & 7.0 & $0.80\pm 0.03$ & $0.62\pm 0.03$ && 0.84 &	     && 0.87 & 0.71 \\
  5.5 & 7.0 & $0.41\pm 0.02$ & $0.62\pm 0.03$ && 0.51 & 0.92 && 0.46 & 0.71 \\
  \hline\hline
  \end{tabular}
\end{table*}

\begin{table*}
  \caption{Comparison of the present $R(N_e)$ and $G(T_e)$ line ratios in \ion{Si}{xiii} with those measured in a tokamak experiment and previously calculated \citep{kee89}.$^b$ \label{tokamak_si}}
  \centering
  \small
  \renewcommand
  {\arraystretch}{1.3}
  \begin{tabular}{cccccccccc}
  \hline\hline
  & &	\multicolumn{2}{c}{Present work} && \multicolumn{2}{c}{Expt$^b$} && \multicolumn{2}{c}{Theory$^b$}\\
  \cline{3-4} \cline{6-7}\cline{9-10}
  $N_e(10^{13}$\,cm$^{-3}$) & $T_e(10^2$\,eV) & $R(N_e)$ & $G(T_e)$ && $R(N_e)$ & $G(T_e)$ && $R(N_e)$ & $G(T_e)$ \\
  \hline
  1.2 & 8.00 & $1.86\pm 0.09$ & $0.66\pm 0.04$ && 1.94 & 0.73 && 2.03 & 0.73 \\
  1.5 & 7.50 & $1.86\pm 0.09$ & $0.69\pm 0.04$ && 1.68 & 0.74 && 1.92 & 0.74 \\
  3.0 & 7.50 & $1.50\pm 0.07$ & $0.68\pm 0.04$ && 1.42 & 0.71 && 1.63 & 0.74 \\
  6.0 & 6.50 & $1.16\pm 0.05$ & $0.74\pm 0.04$ && 1.27 & 0.84 && 1.26 & 0.77 \\
  9.0 & 7.00 & $0.95\pm 0.04$ & $0.72\pm 0.04$ && 1.13 & 0.92 && 1.01 & 0.76 \\
  \hline\hline
  \end{tabular}
\end{table*}					

\section{Tokamak benchmark}
\label{tokamak}

Most appropriately, the present $R(N_e)$ and $G(T_e)$ line ratios for \ion{Mg}{xi} and \ion{Si}{xiii} can also be compared with tokamak measurements at electron densities and temperatures similar to those of coronal plasmas and with an earlier simulation that attempted to reproduce them \citep{kee89, kee91}. A key aspect of this benchmark is that the electron density and temperature were measured with alternative experimental techniques, namely microwave interferometry and laser--electron scattering, and therefore, a comparison of the spectroscopic line ratios with theory would lead to validations of the experimental accuracy (stated at $\sim 10\%$ in $N_e$ and $T_e$), the spectral computational model and the reliability of the atomic data. A notable difference between our spectral model and that used by \citet{kee89, kee91} is their inclusion of radiative and dielectronic recombination of the H-like ion and, in the case of \ion{Si}{xiii}, also of the inner-shell ionization of the Li-like ion. Regarding the accuracy of the effective collision strengths used in these previous calculations, the data sets were taken from several sources; but specifically for \ion{Si}{xiii}, \citet{kee87} provide analytic expressions that can be compared with the values of \citet{agg10} adopted in the present work. The agreement has been found to be better than 10\% in the temperature range of interest.
								
In Tables~\ref{tokamak_mg}--\ref{tokamak_si} the present $R(N_e)$ and $G(T_e)$ line ratios in \ion{Mg}{xi} and \ion{Si}{xiii} are respectively compared with the tokamak measurements and previous theoretical values reported by \citet{kee89, kee91}. In general, the present line ratios are somewhat lower than previous theory but by not more than 13\% for \ion{Mg}{xi} and 10\% for \ion{Si}{xiii}; such differences, in our opinion, are mostly due to electron recombination that has been excluded in our model. The comparison with experiment yields larger differences: for \ion{Mg}{xi}, differences in $R(N_e)$ are within 20\% relative to the present values and within 10\% to the computed data of \citet{kee91}, while those in $G(T_e)$ grow up to 33\% and 23\%, respectively, at the higher densities ($5.5\times 10^{13}$\,cm$^{-3}$); for \ion{Si}{xiii}, differences in $R(N_e)$ are within 15\% in both calculations \citep[present and][]{kee89}, but those in $G(T_e)$ again show differences as large as 22\% and 17\%, respectively, at $N_e=9.0\times 10^{13}$\,cm$^{-3}$.

\begin{table*}
  \caption{\ion{Ne}{ix} target-level key (a full version of this table is available online). \label{EnergyNe}}
  \centering
  \begin{tabular}{lclrrrrr}\hline\hline
Index	&	$n$ & Level	&	$(2S+1)$	&	$L$	&	$\pi$	&	$J$	&	Energy (Ryd)	\\\hline
1	&	1 & $^1{\rm S}_0$	&	1	&	0	&	0	&	0	&	0.000000	\\
2	&	2 & $^3{\rm S}_1$	&	3	&	0	&	0	&	1	&	66.521948	\\
3	&	2 & $^3{\rm P}_0$	&	3	&	1	&	1	&	0	&	67.235106	\\
4	&	2 & $^3{\rm P}_1$	&	3	&	1	&	1	&	1	&	67.237839	\\
5	&	2 & $^3{\rm P}_2$	&	3	&	1	&	1	&	2	&	67.251964	\\
6	&	2 & $^1{\rm S}_0$	&	1	&	0	&	0	&	0	&	67.275930	\\
7	&	2 & $^1{\rm P}_1$	&	1	&	1	&	1	&	1	&	67.766921	\\
\vdots	&	\vdots	&	\vdots	&	\vdots	&	\vdots	&	\vdots	&	\vdots	&	\vdots\\\hline\hline
  \end{tabular}
\end{table*}

\begin{table*}
  \caption{Effective collision strengths for \ion{Ne}{ix} (a full version of this table is available online).\label{UpsilonNe}}
  \centering
  \begin{tabular}{ccccccccccc}\hline\hline
	&		& \multicolumn{8}{c}{$\log(T_e/K)$}\\ \cline{3-11}
k	&	i	&	4.40	&	4.60	&	4.80	&	5.00	&	5.20	&	5.40	&	5.60	& $\cdots$ &	7.20	\\\hline
2	&	1	&	1.813E$-$02	&	1.947E$-$02	&	1.894E$-$02	&	1.669E$-$02	&	1.369E$-$02	&	1.084E$-$02	&	8.573E$-$03	& $\cdots$ &	1.497E$-$03	\\
3	&	1	&	4.371E$-$03	&	3.460E$-$03	&	2.833E$-$03	&	2.390E$-$03	&	2.090E$-$03	&	1.909E$-$03	&	1.806E$-$03	& $\cdots$ &	6.473E$-$04	\\
3	&	2	&	5.205E$-$01	&	5.174E$-$01	&	5.180E$-$01	&	5.226E$-$01	&	5.325E$-$01	&	5.492E$-$01	&	5.734E$-$01	& $\cdots$ &	9.972E$-$01	\\
4	&	1	&	1.615E$-$02	&	1.228E$-$02	&	9.689E$-$03	&	7.915E$-$03	&	6.737E$-$03	&	6.021E$-$03	&	5.606E$-$03	& $\cdots$ &	1.964E$-$03	\\
4	&	2	&	1.525E$+$00	&	1.529E$+$00	&	1.539E$+$00	&	1.558E$+$00	&	1.591E$+$00	&	1.643E$+$00	&	1.716E$+$00	& $\cdots$ &	2.991E$+$00	\\
4	&	3	&	9.303E$-$02	&	9.073E$-$02	&	8.787E$-$02	&	8.484E$-$02	&	8.361E$-$02	&	8.565E$-$02	&	8.925E$-$02	& $\cdots$ &	2.466E$-$02	\\
\vdots	&\vdots	&\vdots	&\vdots	&\vdots	&\vdots	&\vdots	&\vdots	&\vdots	&$\cdots$	&\vdots	\\\hline\hline
  \end{tabular}
\end{table*}	

\section{Conclusions}
\label{conclusions}

\begin{enumerate}
  \item Regarding the accuracy of the effective collision strengths required to model the He-like triplet in ions with atomic number $Z\leq 14$, it has been shown that the data sets computed by \citet{zha87} for transitions between levels with $n\leq 2$ and those by \citet{agg08, agg10, agg12} and \citet{agg09, agg11} for transitions with $n\leq 3$ are highly reliable (much better than 10\%). On the other hand, due to gross discrepancies for some transitions, data sets by \citet{del02} for \ion{O}{vii} and by \citet{che06} for \ion{Ne}{ix} should be treated with caution. For \ion{Ne}{ix}, we strongly recommend the effective collision strengths for $n\leq 3$ transitions obtained in the present work that are tabulated in Tables~8--9. In spite of the excellent agreement between the present {\sc bprm} effective collision strengths for transitions between $n\leq 3$ levels in He-like ions and those computed with {\sc darc}, which may indeed stand as a benchmark, there are still some unexplained large discrepancies in the effective collision strengths for E1 transitions between degenerate or quasi-degenerate levels with initial and final principal quantum numbers $n_i=n_k=4$ and $n_i=n_k=5$, which would require further study before more precise rankings can be assigned.

  \item Uncertainty intervals as propagated from the atomic data errors have been determined for the first time for the $R(N_e)$ and $G(T_e)$ line ratios assuming both MB and $\kappa$ electron-energy distributions. For the $R(T_e)$ density diagnostic, it is shown that these uncertainty bands are mainly conspicuous in the low-density region and, therefore, do not affect the diagnostic effectiveness, but they limit the gauging of the line-ratio lowering caused by radiative excitation of the $2\,^3{\rm S}_1$ level. For the $G(T_e)$ line ratio, they shorten the effective temperature range of the diagnostic. They also make the density and temperature diagnostics computed with atomic models containing levels with $n>3$ indistinguishable from those constructed with $n\leq 3$.

  \item $\kappa$-averaged effective collision strengths at $\kappa\lesssim 10$ are enhanced  with respect to the MB values for excitation, in particular for the resonance ($r$) transition, while those for de-excitation are hardly modified. The $\kappa$ distribution significantly lowers the $G(T)$ line ratio at low temperatures although the uncertainty bands due to the atomic data errors blot out the evidence at the higher temperatures.

  \item Our survey of stellar coronal observations of the He-like triplet in ions with atomic number $Z\leq 14$ and the subsequent spectral reduction with the {\sc xspec} code confirm previous observational estimates of the $R(N_e)$ and $G(T_e)$ line ratios, although in some cases we have managed to improve somewhat the fit statistics. It is found that the observational error bars are far too large to be able to discern the electron-energy distribution type and the definite plasma coexistence of the studied ionic species. However, the diagnostic maps of Fig.~\ref{benchmark} seem to broadly indicate at least two plasma components: a high-temperature ($T_e\gtrsim 6$~MK) region containing \ion{Ne}{ix}, \ion{Mg}{xi} and \ion{Si}{xiii}; and a low temperature counterpart ($T_e\lesssim 2$~MK) where \ion{C}{v}, \ion{N}{vi} and \ion{O}{vii} coexist.

   \item With the exception of massive O stars with strong winds such as $\zeta$ Puppis, where line-profile inconsistencies can be apparently solved by invoking resonant scattering \citep{leu07}, and the coronae of the active II Peg and IM Peg binaries \citep{tes04b}, there is scanty evidence of this process in the stellar coronae sampled in this work \citep{nes03b}. The latter authors even consider that manifestation of resonant scattering in the solar corona remains controversial; therefore, this process has not been included in the present plasma model. On the other hand, it may be argued that the somewhat high observed line ratios ($1.2\leq G(T_e)\leq 1.7$) of \ion{C}{v} and \ion{N}{vi} depicted in Fig.~\ref{benchmark} could be due to resonant scattering. We have looked into this possibility by assigning the resonance line an attenuated radiative rate of the form $A_{\rm rs}=A_{w}\exp(-\tau)$, and increasing the optical depth $\tau$ to shift up the theoretical locus closer to the observed points. It is found that this only occurs in \ion{C}{v} at unrealistic values of $\tau\gg 1$.

   \item As previously discussed by \citet{kee89, kee91}, the tokamak benchmark brings out the relative importance of including, in the spectral model of the He-like triplet, the radiative and dielectronic recombination of the H-like ion and the inner-shell ionization of the Li-like ion, even in a collisionally dominated plasma such as a stellar corona. Although these contributions are mainly expected to be conspicuous in the higher-$Z$ ions (e.g. \ion{Mg}{xi} and \ion{Si}{xiii}), \citet{meh15} have recently found evidence that the \ion{O}{vii} triplet components are significantly modified by \ion{O}{vi} line absorption in optically thin plasmas, thus leading to unreliable density diagnostics. Taking into consideration that in many previous coronal plasma models \citep{nes03b} and in the present one these contributions have been neglected, we firmly recommend their inclusion to ensure accurate diagnosing. However, since we were mainly interested in a benchmark of the effective collision strengths and their propagated effects on the line-ratio diagnostics, the construction of more refined models with additional impinging variables would depart from our initial motivation and line of action. It would actually involve further lengthy evaluations of a variety of atomic data sets that lie outside the scope of the initial project. Nonetheless, they could indeed be tackled in a follow-up venture.
\end{enumerate}

\section*{Acknowledgements}

This work has been funded by IVIC. We are indebted to Dr. K.~M. Aggarwal (Queen's University, Belfast, UK) for kindly giving us access to the {\sc darc} collision-strength energy tabulations for the \ion{C}{v}, \ion{N}{vi}, \ion{O}{vii}, \ion{Mg}{xi} and \ion{Si}{xiii} ionic systems; and to Drs. Randall Smith and Adam Foster both from the Harvard--Smithsonian Center for Astrophysics, Cambridge, MA, USA, for allowing us to use a beta version of the {\tt PyAtomDB} Python utility to manipulate the {\sc atomdb} atomic database. Private communications with Drs. A.~K. Pradhan (Ohio State University, Columbus, OH, USA), F. Delahaye (Observatoire de Paris, Meudon, France) and G.-X. Chen are acknowledged. We are grateful to the referee, Professor Francis Keenan, for suggesting the tokamak benchmark that became an integral part of the paper and for other comments that led to improvements of its scientific content.




\bibliographystyle{mnras}





\bsp	
\label{lastpage}
\end{document}